\documentclass[12pt,a4paper]{article}
\usepackage{comment}
\usepackage{amssymb}
\usepackage{amsmath}
\usepackage{amsfonts}
\usepackage{enumerate}
\usepackage{float}
\usepackage{natbib}
\usepackage{verbatim}
\usepackage{graphicx}
\usepackage{caption}
\usepackage{lscape}
\usepackage[autostyle]{csquotes}
\usepackage{bbm}
\usepackage[dvipsnames]{xcolor}
\usepackage[a4paper,bindingoffset=0.2in,left=1in,right=1in,top=1in,bottom=1in,footskip=.5in]{geometry}
\usepackage{mathrsfs} 
\usepackage{xfrac}
\DeclareMathOperator{\plim}{plim}
\usepackage{multirow}
\usepackage{longtable}
\usepackage{float}

\usepackage{esvect}
\usepackage[T1]{fontenc}
\usepackage{amsmath}
\usepackage{setspace}
\usepackage[utf8]{inputenc}
\usepackage{xcolor}
\usepackage{mathrsfs} 
\usepackage[autostyle]{csquotes}

\restylefloat{table}

\graphicspath{ {images/} }   

\DeclareMathAlphabet{\pazocal}{OMS}{zplm}{m}{n}

\begin{document}
\title{Consistent Specification Test of the Quantile Autoregression\thanks{%
I am greatly indebted to Valentina Corradi and Vasco Gabriel for their constant guidance throughout this project. I would also like to thank Joao Santos Silva, Federico Martellosio, Ekaterina Oparina, Ayden Higgins, Francesco Vitiello, Khashayar Rahimi and the participants of the University of Surrey Econometric Workshop, the RES Symposium of Junior Researchers 2019, the IAAE 2019 and the ESEM 2019 conference for helpful comments and discussions.The author also thanks the South East Network for Social Sciences for financial support.}}
\author{Anthoulla Phella{\small \thanks{\textit{E-mail address}:
anthoulla.phella@glasgow.ac.uk}} \\
University of Glasgow}
\date{\today}
\maketitle

\begin{abstract}
This paper proposes a test for the joint hypothesis of correct dynamic specification and no omitted latent factors for the Quantile Autoregression. If the composite null is rejected we proceed to disentangle the cause of rejection, i.e., dynamic misspecification or an omitted variable. We establish the asymptotic distribution of the test statistics under fairly weak conditions and show that factor estimation error is negligible. A Monte Carlo study shows that the suggested tests have good finite sample properties. Finally, we undertake an empirical illustration of modelling growth and inflation in the United Kingdom, where we find evidence that factor augmented models are correctly specified in contrast with their non-augmented counterparts when it comes to GDP growth, while also exploring the asymmetric behaviour of the growth and inflation distribution. 


\end{abstract}

\vspace{2cm}

\setcounter{page}{1}\newpage
\section{Introduction}
Quantile regression (QR) enables the analysis of a continuous range of conditional quantile functions, providing a more complete picture of the conditional dependence structure of the variables examined, rather than a single measure of conditional location. As the awareness of the importance of data heterogeneity increases, quantile regression has become even more relevant \citep{koenker_quantile_2017}. At the same time, the recent availability of large datasets has generated interest in models with many possible predictors. Inference methods overcoming the curse of dimensionality have become increasingly popular and factors in particular, have been proven useful in overcoming the limited information bias. This arises from the fact that the information set of decision makers is sufficiently larger than the information set captured by conventional empirical models. \\

The intersection of latent factors with quantile regression models is fairly recent.  \cite{ando_quantile_2011} have considered a quantile regression model with factor-augmented predictors, whose effect is allowed to vary across the different quantiles. Their study models the quantile structure in a cross-sectional context.  More recently, \cite{ando_quantile_2020} introduced a new procedure for analysing the quantile co-movement of a large number of time series based on a large scale panel data model with factor structures. In their study the latent factors are allowed to vary across the different quantiles of the variables from which they are extracted and as such their model is a quantile factor model. Similarly, \citet{chen_quantile_2019} estimate scale-shifting factors and quantile dependent loadings, thus factors may shift characteristics (moments or quantiles) of the distribution of the set of directly observable measures, other than its mean, and factor loadings are allowed to vary across the distributional characteristics of each variable. We add in this relevant literature by using mean shifting factors, that is factors that are only allowed to alter the location of the observable measures, as a method of dimension reduction and use these latent factors as additional regressors in quantile autoregressive models. \\

Meanwhile, in the empirical literature, the majority of the work has examined whether univariate regression models should be augmented with factors, in estimates and forecasts of the conditional mean (\citealp{stock_forecasting_2002, bernanke_measuring_2005}). However, point estimates and forecasts for the conditional mean of macroeconomic variables, like growth and inflation, ignore the risks around this central estimates \citep{adrian_vulnerable_2019}. Furthermore, over the recent years, the tendency of policy-makers to focus more on the downside risk of GDP growth demonstrates the need to  analyse the conditional quantiles of GDP growth and examine how they should be modelled individually. Similarly,  as supported by \citet{levin_is_2002} and \citet{angeloni_new_2006}, the inflation rate is one of the most important variables, due to its dominant role in many macroeconomic models and as argued by, inter alia, \citet{henry_is_2004}, the dynamic behaviour of the inflation rate has a number of economic implications. In order to correct for this omission, we trace the conditional distributions of GDP growth and CPI inflation in the United Kingdom by estimating several conditional quantiles. \\

In this paper therefore, our contribution to the existing literature is twofold. Firstly, we allow for the interaction of mean shifting factors, summarising a larger information set, with quantile autoregressive models and furthermore, as a result, obtain a trace of the conditional distribution of a variable of interest. The latter enables us to assess the upside and downside risks present for the dependent variable. We focus on the GDP growth and CPI inflation in the United Kingdom and find that quantile autoregressive models for GDP growth should include latent factors, as a way to summarise multiple macroeconomic variables. Such factors have non-uniform effects on different tails of the growth distribution. However, these latent factors do not carry relevant information for modelling the inflation rate distribution. We also find evidence of a decrease in the asymmetry of UK GDP growth following the recent financial crisis and significant time variation in downside risk. Meanwhile, in the case of inflation upside risk has varied significantly over the years despite a rather stable central tendency.  \\

Nevertheless, for any post estimation inference to be valid, the correct specification of the empirical model used needs to be assessed. Therefore, we also propose a test for the joint hypothesis of correct dynamic specification and no omitted latent factor for the quantile autoregression (QAR), as that is outlined in \cite{koenker_quantile_2006}. The test we suggest is related to the conditional moment tests of \citet{bierens_consistent_1982, bierens_consistent_1990},  as well as the specification test of parametric quantile models of \citet{escanciano_specification_2010}\footnote{Though similar, the work by \citet{escanciano_specification_2010}  cannot account for the factor estimation error present in our context due to the use of unobservable factors that need to be estimated a priori.}. In practice, we propose two tests: the first one can be used to determine if the quantile autoregression is correctly specified, conditional on all available past information of the dependent variable and latent factors. In cases where the null hypothesis fails to reject when this test is applied, we have evidence that the QAR is correctly specified. If however, the null hypothesis is rejected, we perform a second test in order to determine whether the misspecification arises because latent factors are an omitted variable or because the model is dynamically misspecified. The second test therefore involves testing the null hypothesis of the Factor-Augmented QAR (FA-QAR) being correctly specified. Valid asymptotic critical values are obtained via a bootstrap procedure based on resampling functions of the entire history of available information as in \citet{corradi_information_2009}.\\

The remainder of the paper is organised as follows. In Section 2, we outline the framework and describe the testing procedure in order to examine whether a Quantile Autoregression should be augmented with latent factors. We define the test statistics and study the asymptotic properties of the suggested statistics.  Section 3 demonstrates some finite sample performance results based on a limited Monte Carlo simulation. In Section 4, we examine the distributions of UK growth and inflation, based on the use of a large-scale macroeconomic dataset and discuss the empirical findings. Concluding remarks are given in Section 5 and information regarding proofs is referred to an appendix.\\

\vspace{1cm}

\section{The Framework}
\subsection{Test Statistics}
We begin by outlining the factor model used in the sequel. Let
\begin{align}
X_t = \Lambda_t F_t + e_t  \quad \label{factor extraction}
\end{align}
where  $X_t$ is an $N \times 1$ vector of observable variables characterising the economy, $\Lambda_t$ is an $N \times k$ matrix of factor loadings, $F_t$ is a $k \times 1$ vector of the k latent common factors and $e_t$ is an $N \times 1$ vector of idiosyncratic disturbances. The errors are allowed to be both serially and (weakly) cross sectionally correlated.\\

As in \citet{stock_forecasting_2002}, factors are extracted via the principle components approach and the estimated factors and estimated factor loadings are defined as:
\begin{align*}
(\hat{F},\hat{\Lambda})=\arg \min_{F,\Lambda} \frac{1}{NT}\sum_{i=1}^N \sum_{t=1}^T (X_{it}-\Lambda_{i}F_{t})
\end{align*}

The resulting principal components estimator of $F$ is then $\hat{F}=\frac{X'\hat{\Lambda}}{N}$, where $\hat{\Lambda}$ is set equal to the eigenvectors of $X'X$ corresponding to its $k$ largest eigenvalues. In the remainder of this paper the number of factors $k$ would remain fixed and can be estimated using the information criteria outlined in \citet{bai_determining_2002} who take into account the sample size both in the cross-section and time-series dimensions.\\ 

Suppose we observe a real-valued dependent variable $y_t$ and a high-dimensional information vector $X_t$. Due to the curse of dimensionality we wish to reduce the dimension of $X_t$ with the use of factors, as a way to summarise all the available information. The desirable information vector is therefore $I_{t-1}=(Y'_{t-1}, F'_{t-1}) \in \Re^d$, $d=(p+1)+k$, where $F_{t-1}=(F_{1,t-1},...,F_{k,t-1}) \in \Re^k, k\in \aleph$, is the vector of  latent factors and $Y_{t-1}=(1, y_{t-1},...,y_{t-p}) \in \Re^{p+1}$, where $A'$ denotes the transpose of $A$. In practice, the vector $F_{t-1}$ is unobservable, therefore, we replace the infeasible information set $I_{t-1}$ with the feasible information vector $\hat{I}_{t-1}=(Y'_{t-1}, \hat{F}'_{t-1}) \in \Re^d$, $d=(p+1)+k$, where $\hat{F}_{t-1}=(\hat{F}_{1,t-1},...,\hat{F}_{k,t-1}) \in \Re^k, k\in \aleph$, is the vector of  estimated factors from the panel data $X_{i,t-1}$. In the remainder of this paper we assume that the time series process $\lbrace(Y_t, F_{t-1}')':t=0,\pm 1,\pm 2,...\rbrace$, defined on the probability space $(\Omega, \mathcal{A})$ is strictly stationary and ergodic.\\

Under the assumption that the conditional distribution of $Y_t$ given $I_{t-1}$ is continuous, we can then define the $\tau^{th}$ conditional quantile of $Y_t$ given $I_{t-1}$ as the measurable function $q_{\tau}$ satisfying the conditional restriction
\begin{align}
P(y_t \leq q_{\tau}(I_{t-1}) \mid I_{t-1})=\tau, \text{ almost surely (a.s).}\quad \label{Conditional Quantile}
\end{align}\\
We use the Quantile Autoregression (QAR) model of order p as that is set out in  \citet{koenker_quantile_2006} so that the variable of interest can then be characterised by the following equation
\begin{align}
y_t = \theta_0(u_t)+\theta_1(u_t)y_{t-1} +...+\theta_p (u_t) y_{t-p},  \label{inflation}
\end{align}
where ${u_t}$ is a sequence of i.i.d. standard uniform random variable, while $\theta_i(u_t)$ are unknown functions $[0,1] \rightarrow \Re$.  Provided that the right hand side of (\ref{inflation}) is monotone increasing in $u_t$, the $\tau^{th}$ conditional quantile function of $y_t$ takes the following form,
\begin{align}
m(Y_{t-1}, \theta(\tau))&=\theta_0(\tau)+\theta_1 (\tau)y_{t-1} +...+\theta_p (\tau) y_{t-p} \notag  \\
&=Y_{t-1} \theta(\tau). \quad  \label{Linear QAR} 
\end{align}\\
Our objective is to test whether factors are an omitted variable in a quantile autoregressive model for $y_t$, while at the same time controlling for dynamic misspecification. In the sequel, we are testing the following hypothesis:
\begin{align}
&H_{0,1}: \quad E[\mathbbm{1}(y_t-Y_{t-1}\theta(\tau)) \leq 0)-\tau|I_{t-1}]=0  \quad \quad \quad \text{a.s. for some $\theta \in \mathcal{B}$ and for all $\tau \in \mathcal{T}$} \label{Null Hypothesis 1}
\end{align} 
against its respective alternative
\begin{align}
&H_{A,1}: \quad Pr\{E[\mathbbm{1}(y_t-Y_{t-1}\theta(\tau)) \leq 0)-\tau | I_{t-1}]=0\}<1, \quad \quad \quad \quad \quad \quad \quad \quad \quad \quad \quad \quad \quad \quad \quad   \label{Alternative Hypothesis 1}
\end{align}
where $\mathcal{B}$ is a family of uniformly bounded functions from $\mathcal{T}$ to $\Theta$.\\

The null hypothesis states that if the specification is correct, the probability that the observed value of $y_t$ falls below the estimated quantile should, on average, equal the nominal quantile level of interest ($\tau$), with probability one. Hypothesis $H_{0,1}$ is the joint hypothesis that the QAR model is not dynamically misspecified and factors are not an omitted variable. If $H_{0,1}$ did not include such latent factors, the null hypothesis would simplify to the null hypothesis of the correct specification of parametric dynamic quantile models,  tested in \citet{escanciano_specification_2010}, where the conditioning information vector includes only directly observable variables.\\

Testing for the null hypotheses $H_{0,1}$ is a challenging problem, since it involves an infinite number of conditional moments indexed by $\tau \in \mathcal{T}$, where $\mathcal{T}$ is a compact set comprising of the range of quantiles of interest ($\mathcal{T}\subset(0,1)$). Therefore, following \citet{bierens_consistent_1990} we can characterise $H_{0,1}$ by the infinite number of unconditional moment restrictions:
\begin{align}
&E\{[\mathbbm{1}(y_t-Y_{t-1}\theta(\tau))\leq 0)-\tau]exp(\xi'\Phi(I_{t-1}))\}=0
\end{align}
where $\Phi$ is an arbitrary Borel Measurable bounded one-to-one mapping from $\Re^d$ to $\Re^d$ and $\xi \in \Re^d$ is a  vector of weights. Conditioning on $I_{t-1}$ is equivalent to conditioning on the bounded vector $\Phi(I_{t-1})$, for $I_{t-1}$ and $\Phi(I_{t-1})$ generate the same Borel field. Furthermore, we wish for the weight attached to past observations to decrease over time, thus we define the weighting vector $\xi$ as in \citet{de_jong_bierens_1996}. Let therefore, $exp(\xi' \Phi(Z_t))=exp(\sum_{j=1}^{t-1} \xi'_j \Phi(Z_{t-j}))$ and $\Xi=\lbrace \xi_j: a_j\leq \xi_j \leq  \gamma_j, j=1,2; \vert a_j \vert, \vert \gamma_j \vert \leq \Gamma j^{-\kappa}, \kappa \geq 2\rbrace$.\\

Given therefore a sample $\{(y_t, \hat{I}'_{t-1})': 1 \leq t \leq T\}$ and an estimated parameter value $\hat{\theta}(\tau)$ we consider the following quantile empirical process: \begin{align}
S_{1,T}(\xi,\hat{\theta}_T(\tau))&:= T^{-\frac{1}{2}}\sum_{t=1}^{T} [\mathbbm{1}(y_t-Y_{t-1}\hat{\theta}_T(\tau)\leq 0)-\tau]exp(\xi'\Phi(\hat{I}_{t-1}))\label{Emprirical Process}
\end{align}
for the Quantile Autoregression Estimator (QARE), proposed by a \citet{koenker_quantile_2006}), defined as 
\begin{align}
\hat{\theta}_T(\tau) = \arg\min_{\theta \in \mathbb{R}^{p+1}} \sum_{t=1}^{T} \rho_{\tau} (y_t-Y_{t-1}\theta) \label{Estimator}
\end{align} 
where $\rho_{\tau}(u)=u(\tau-\mathbbm{1}(u<0))$ is the \enquote{tick} loss function.\\

The null hypothesis holds when the process $S_{1,T}(\xi,\hat{\theta}_T(\tau))$ is close to zero for almost all $(\xi',\tau)' \in \Re^d \times \mathcal{T}$, and thus  the test statistic is based on a distance from a standardised sample analogue of $E\{[\mathbbm{1}(y_t-Y_{t-1}\hat{\theta}_T(\tau)\leq 0)-\tau]exp(\xi'\Phi(\hat{I}_{t-1}))\}$ to zero. Some popular norms we could consider are the following:
\begin{itemize}
\item \label{CvM}{Cramer-von-Mises $\Rightarrow CvM_{j,T}:=\int_{\mathcal{T}}\int_{\mathcal{Y}}|S_{1,T}(\xi,\hat{\theta}_T(\tau))|^2d\Phi_1(\xi)d\Phi_2(\tau)$}
\item \label{KS}{Kolmogorov-Smirnov $\Rightarrow KS_{j,T}:=sup_{\tau \in \mathcal{T}}\int_{\mathcal{Y}} |S_{1,T}(\xi,\hat{\theta}_T(\tau))|^2d\Phi_1(\xi)$}
\end{itemize}
where $\Phi_1$ and $\Phi_2$ are some integrating measures on $\mathcal{Y}$ and  $\mathcal{T}$ respectively and $\mathcal{Y}$ is a generic compact subset of $\Re^d $ containing the origin.\\

The test we propose therefore rejects $H_0$ for \enquote{large} values of such functionals. If $H_{0,1}$ is not rejected then one can conclude that the QAR model is correctly specified with no omitted variables and thus can proceed with inference. If however, $H_{0,1}$ is rejected we still need to ascertain the source of the rejection. A logical next step would therefore be to augment the quantile autoregression model with the feasible estimated factors. We therefore proceed to test the following null hypothesis: 
\begin{align}
&H_{0,2}: \quad E[\mathbbm{1}(y_t-Y_{t-1}\theta_1(\tau)-F_{t-1}\theta_2(\tau) \leq 0)-\tau|I_{t-1}]=0  \quad \text{a.s. for some $\theta \in \mathcal{B}$ and for all $\tau \in \mathcal{T}$}\label{Null Hypothesis 2} 
\end{align} 
against its negation
\begin{align}
&H_{A,2}: \quad Pr\{E[\mathbbm{1}(y_t-Y_{t-1}\theta_1(\tau)-F_{t-1}\theta_2(\tau) \leq 0)-\tau | I_{t-1}]=0\}<1 \quad \quad \quad \quad \quad \quad \quad \quad  \quad \quad \quad\label{Alternative Hypothesis 2}
\end{align}
In this case, the null hypothesis $H_{0,2}$ states that the Factor-augmented QAR (FA-QAR) is correctly specified. It follows that the test statistic of interest for $H_{0,2}$ will be based on the quantile empirical process $S_{2,T}(\xi,\hat{\theta}_T(\tau)):= T^{-\frac{1}{2}}\sum_{t=1}^{T} [\mathbbm{1}(y_t-Y_{t-1}\hat{\theta}_{1,T}(\tau)-\hat{F}_{t-1}\hat{\theta}_{2,T}(\tau)\leq 0)-\tau]exp(\xi'\Phi(\hat{I}_{t-1}))$. 
In this instance the factor estimation error not only appears in the conditioning information vector but also influences the indicator function element of the statistic.
If the null hypothesis fails to reject then the FAQAR is correctly specified, hence factors were originally an omitted variable. If on the other hand, $H_{0,2}$ is also rejected then we have evidence that the linear specification of the model may not be appropriate and a non-parametric approach may be more suitable, however this is beyond the scope of this paper.\\

\vspace{0.5cm}
\subsection{Assumptions}
Let $\Vert A \Vert =[tr(A'A)^{\frac{1}{2}}]$ denote the norm of matrix A. Throughout, we let $F_t$ be the $k \times 1$ vector of true factors and $\lambda_i$ be the true loadings, with $F$ and $\Lambda$ being the corresponding matrices. The relevant assumptions for the latent factors are those used in \citet{bai_inferential_2003}  to derive the limiting distributions of the estimated factors, factor loadings and common components and are provided explicitly in the Appendix as Assumptions A-F. We rely on those same assumptions, to demonstrate that factor estimation error does not influence our test statistic.\\

To derive the asymptotic results of the quantile empirical process, we need to further consider the following assumptions. Let, for each $t \in \mathbb{Z}$, $\mathcal{F}_t=\sigma (I_t', I_{t-1}',\ldots)$ be the $\sigma$-field generated by the information set obtained up to time $t$. Define also the family of conditional distributions $F_b(y)\mathrel{\mathop:}= P(Y_t \leq y \vert I_{t-1}=b)$ . Let $f_b$ be the density function of the cumulative distribution function (cdf) $F_b$. In particular, $f_{I_{t-1}}(y)$ denotes the density of $Y_t$ given $I_{t-1}$, evaluated at $y$. Also, the family $\mathcal{B}$, in which the parameter $\theta$ takes values, is endowed with the sup norm\footnote{i.e. $\Vert \theta\Vert_{\mathcal{B}}=\sup_{\tau \in \mathcal{T}} \vert \theta(\tau) \vert$.}. Lastly, given that similar assumptions are needed both when testing the QAR model under $H_{0,1}$ and FA-QAR model under $H_{0,2}$,  let $m_j(I_{t-1},\theta(\tau))$ be the conditional quantile function under consideration and $H_{0,j}$ the null hypothesis to be tested. Thus,  $m_1(I_{t-1},\theta(\tau))=Y_{t-1} \theta(\tau)$ is the conditional quantile function in the QAR case and $m_2(I_{t-1},\theta(\tau))=Y_{t-1}\theta_1(\tau)+F_{t-1}\theta_2(\tau)$ in the FA-QAR case. \\

\vspace{0.25cm}

\textbf{Assumption G: Time series Model Checks}

\quad \quad 1. $\lbrace(Y_t, F_t')':t=0, \pm 1, \pm2,\ldots\rbrace$ is a strictly stationary and ergodic process. Under $H_{0,j}$, $\lbrace \mathbbm{1}(y_t-m_j(I_{t-1},\theta(\tau))\leq 0)-\tau,\mathcal{F}_t\rbrace$ is a martingale difference sequence for all $\tau \in \mathcal{T}$.\\

\quad \quad 2.  $m_j(I_{t-1},\theta(\tau))$ is non-decreasing in $\tau$ a.s.\\

\quad \quad 3. The family of distributions functions $\lbrace F_b, b \in \Re^d \rbrace$ has Lebesque measures $\lbrace f_b, b \in \Re^d \rbrace$ that are uniformly bounded away from zero for the quantiles of interest.\\

\vspace{0.25cm}

\textbf{Assumption H: Class of functions}\\
For each general $\theta_1 \in \mathcal{B}$,\\

\quad \quad 1. There exists a vector of functions $g_{t-1}: \Theta\rightarrow \Re^q$ such that $g_{t-1}(\theta_1(\tau))$ is $\mathcal{F}_{t-1}$-measurable for each $t \in \mathcal{Z}$, and satisfies for all $k<\infty$,
\begin{align*}
\sup_{1 \leq t \leq n, \Vert \theta_1-\theta_2 \Vert_{\mathcal{B}\leq kT^{-\frac{1}{2}}}} T^{\frac{1}{2} }\Vert m_j(I_{t-1},\theta_2)-m_j(I_{t-1},\theta_1)-(\theta_2-\theta_1)'g_{t-1}(\theta_1) \Vert_{\mathcal{B}}=o_p(1).
\end{align*}\\

\quad \quad 2. For all sufficiently small $\delta>0$,
\begin{align*}
&E \Big[  \sup_{\Vert \theta_1-\theta_2\Vert _{\mathcal{B}\leq \delta}} \vert  \mathbbm{1}(y_t-m_j(I_{t-1},\theta_1(\tau))\leq 0) -\mathbbm{1}(y_t-m_j(I_{t-1},\theta_2(\tau))\leq 0) \vert \Big] \leq C\delta, \quad \text{for all $\tau \in \mathcal{T}$, and}\\
\\
&E \Big[ \sup_{\vert \tau_1-\tau_2\vert \leq \delta} \vert  m_j(I_{t-1},\theta_1(\tau_1))-m_j(I_{t-1},\theta_1(\tau_2))\vert \Big] \leq C\delta.
\end{align*}\\

\quad \quad 3. Uniformly in $\tau \in \mathcal{T}$, $E \vert g_{t-1}(\theta_1(\tau)) \vert^2 <\infty$, and uniformly in $(\xi, \tau) \in \mathcal{Y} \times \mathcal{T}$,
\begin{align*}\footnotesize{
\vert T^{-1} \sum_{t=1}^{T} g_{t-1}(\theta(\tau)) exp(\xi' \Phi(I_{t-1})) f_{I_{t-1}}(m_j(I_{t-1},\theta_0))- E[g_{t-1}(\theta(\tau)) exp(\xi' \Phi(I_{t-1})) f_{I_{t-1}}(m_j(I_{t-1},\theta_0))] \vert =o_p(1)}.
\end{align*}\\

\vspace{0.25cm}

\textbf{Assumption I: Compactness of the parameter space}\\
The parametric space $\Theta$ is compact in $\Re^p$. The true parameter $\theta(\tau)$ belongs to the interior of $\Theta$ for each $\tau \in \mathcal{T}$ and $\theta \in \mathcal{B}$. The class $\mathcal{B}$ satisfies\\

\quad \quad \quad \quad \quad \quad \quad \quad $\int_0^{\infty} \big( log(N_{[\cdot]}\delta^2, \mathcal{B}, \Vert \cdot \Vert_{\mathcal{B}}) \big)^{\frac{1}{2}}d\delta < \infty$.\\

\vspace{0.25cm}

\textbf{Assumption J: Estimator Consistency}\\
The estimator $\hat{\theta}_T$ satisfies that $P(\hat{\theta}_T \in \mathcal{B})\to 1$ as $T \to \infty$, and the following asymptotic expansion under $H_0$, \\

\quad \quad \quad \quad $Q_T(\tau)=\sqrt{T}(\hat{\theta}_T(\tau)-\theta(\tau))$\\

\quad \quad \quad \quad \quad \quad \quad  $=\frac{M^{-1}}{q(\tau)}T^{-\frac{1}{2}} \sum_{t=1}^T (\mathbbm{1}(y_t-m_j(I_{t-1},\theta(\tau))\leq 0)-\tau)I_{t-1}+o_p(1)$, \quad uniformly in $\tau \in \mathcal{T}$,\\

where $M=E(\textbf{I'I})$ is a positive definite matrix, and $q(\tau)=f_{\epsilon}(F_{\epsilon}^{-1}(\tau))$ is the reciprocal of the sparsity function. Furthermore, the process $Q_T(\tau)$ converges weakly to a zero mean Gaussian process $Q(\tau)$ with covariance function
\begin{align}
\begin{split}
Cov_Q(\tau_1, \tau_2)=\frac{M^{-1}}{q(\tau_1)} \times  L(\theta(\tau_1), \theta(\tau_2)) \times \frac{M^{-1}}{q(\tau_2)} \label{Quantile Process Covariance Function} 
\end{split}
\end{align}\\
where $ L(\theta(\tau_1), \theta(\tau_2)) =\lim_{T\rightarrow \infty}  T^{-1} \sum_{t=1}^T \sum_{s=1}^T  E \Big[ (\mathbbm{1}(y_t-m_j(I_{t-1},\theta(\tau_1))\leq 0)-\tau_1)I_{t-1} \times (\mathbbm{1}(y_s-m_j(I_{s-1},\theta(\tau_2))\leq 0)-\tau_2)I_{s-1} \Big]$.\\

\vspace{0.25cm}

Assumption G1 is standard in time series model checks and is always true in the present context where the information set contains all relevant past history, while G3 is necessary for the asymptotic tightness of the process $S_{j,T}(\xi, \theta_T)$. Assumption H is satisfied for the Linear Quantile Autoregression model under consideration. Sufficient conditions for Assumption I of monotone classes of functions applying to the QAR model can be found in Theorem 2.7.5  in \citet{vaart_weak_2000}.  Meanwhile, the asymptotic normality of the quantile regression process has been established in the literature under a variety of conditions, see for example Theorem 1 in \citet{gutenbrunner_regression_1992}. \\

\vspace{0.5cm}

\subsection{Asymptotic null distribution}
In this subsection we establish the limit distribution of the quantile-marked empirical process $S_{j,T}(\xi,\hat{\theta}_T(\tau))$ under the null hypothesis $H_{0,j}$.\\ 

We begin by showing that the factor estimation error in the feasible information vector is negligible and therefore the proposed statistics converge to the equivalent statistics with the infeasible information vector that includes the true latent factors and a term which converges to zero.\\

\textbf{Lemma 1:} Let Assumptions A-F (shown in Appendix) hold. Then:\\
\begin{align}
S_{j,T}(\xi,\hat{\theta}_T(\tau))&:= T^{-\frac{1}{2}}\sum_{t=1}^{T} [\mathbbm{1}(y_t-m_j(\hat{I}_{t-1},\hat{\theta}(\tau))\leq 0)-\tau]exp(\xi'\Phi(\hat{I}_{t-1})) \notag \\ 
&= T^{-\frac{1}{2}}\sum_{t=1}^{T} [\mathbbm{1}(y_t-m_j(I_{t-1},\hat{\theta}(\tau))\leq 0)-\tau]exp\xi'\Phi(I_{t-1}))+o_p(1)
\end{align}\\

It is immediate to see that, in the first instance when dealing with $S_{1,T}(\xi,\hat{\theta}_T(\tau))$ factor estimation error is only present in the conditioning set and thus only appears in the weighting exponential function, while in the test statistic $S_{2,T}(\xi,\hat{\theta}_T(\tau))$, the factor estimation error is present in both the exponential function and in the indicator function element. Therefore the proof of the statement for $H_{0,1}$ follows from the proof of statement for $H_{0,2}$ show in the Appendix.

\vspace{0.5cm}

We then proceed to state the limiting distribution of the test statistic recognising that in addition to factor estimation error, which has been found to be negligible, there is also parameter estimation error present. Define the function $G(\xi, \theta(\tau))= E[g_{t-1}(\theta(\tau))f_{I_{t-1}}(m_j(I_{t-1},\theta(\tau))exp(\xi' \Phi(I_{t-1}))]$, $\xi \in \Upsilon$, $\tau \in \mathcal{T}$. Also note that under a suitable central limit theorem $S_{j,T}(\xi,\theta_T(\tau))$ converges to a zero mean Gaussian process $S_{j,\infty}(\xi,\theta(\tau))$ with covariance function given by  $Cov_{\infty}(\nu_1, \nu_2)=(\min \lbrace \tau_1,\tau_2 \rbrace-\tau_1\tau_2) E \big[exp((\xi_1-\xi_2)'\Phi(I_0)) \big]$.\\

\textbf{Theorem 1:} Let Assumptions A-J hold. Then, under the null hypothesis $H_{0,j}$,  $S_{j,T}(\xi,\hat{\theta}_T(\tau))\xrightarrow{d} \sup\limits_{\xi \in \Upsilon,\tau \in \mathcal{T}} \vert S_{j}(\xi,\theta(\tau)) \vert$, where $S_{j}(\xi,\theta(\tau)) $ is a zero mean Gaussian process with covariance function 
\begin{align}
\begin{split}
Cov(\nu_1,\nu_2)&=Cov_{\infty}(\nu_1,\nu_2)+G'(\xi_1,\theta(\tau_1))Cov_Q(\tau_1,\tau_2)G(\xi_2,\theta(\tau_2))\\
& \quad \quad -E \big[S_{j,\infty}(\xi_1,\theta(\tau_1))G'(\xi_2,\theta(\tau_2))Q(\tau_2) \big] -E \big[G'(\xi_1,\tau_1)Q(\tau_1)S_{j,\infty}(\xi_2,\theta(\tau_2)) \big]\\ \label{Covariance function}\\
\end{split}
\end{align}
where, $\nu_1=(\xi_1',\tau_1)'$,  $\nu_2=(\xi_2',\tau_2)'$.\\

\vspace{0.5cm}

Details regarding the convergence of the statistic can be found in the Appendix, while the consistency properties of tests based on continuous functionals has been shown in \citet{escanciano_specification_2010}.\\

\vspace{0.5cm}

\subsection{Critical Values Estimation}
Under the null hypothesis, the quantile error $S_{j,T}(\xi,\theta(\tau))$ converges to a Gaussian process with zero mean and a given covariance structure. However, when the estimated parameter $\hat{\theta}_T$ is used in $S_T(\xi,\hat{\theta}_T)$ the parameter estimation error affects its asymptotic properties. Given that the asymptotic null distribution of $S_{j,T}(\xi,\hat{\theta}_T)$ will be dependent on the data generating process and thus is not nuisance parameter free, critical values for the test statistics cannot be tabulated for general cases.\\

Bootstrap methods have been proposed in the literature for quantile regression models (see, e.g., \citet*{hahn_bootstrapping_1995, parzen_resampling_1994}). With respect to quantile regression models with time series, \citet{gregory_smooth_2018} established a smooth tapered block bootstrap procedure. Their work studied the properties of the block bootstrap with smoothing of both data observations via kernel smoothing techniques and data blocks by tapering, which demonstrated the validity of the block bootstrap in dynamic quantile models. At the same time their work extended the validity of the moving block bootstrap \citep{fitzenberger_moving_1998} to quantile regression under weaker conditions than previously considered.\\

In our context, under both the null hypotheses,  $[\mathbbm{1}(y_t-m_j(I_{t-1},\theta(\tau) \leq 0)-\tau]exp(\xi'\Phi(\hat{I}_{t-1})$  is a martingale difference sequence, therefore resampling blocks of length one, as in the iid case preserves the first order validity of the block bootstrap \citep{corradi_information_2009}. In order to achieve higher order refinements, the block bootstrap with an increasing block size would have been necessary, given however that our statistics depend on the nuisance parameters, $\xi$, that are not identified under the null we cannot obtain such refinements. Therefore, in order to preserve the temporal ordering, we proceed to jointly resample ($y_t, Y_{t-1}, \hat{F}_{t-1}, exp(\sum_{j=1}^{t-1} \xi'_j \Phi(I_{t-j}))$) by drawing $T-1$ independent draws. For each bootstrap replication we use the same set of resampled values across $\xi \in \Xi$. \\

The bootstrap analogues of $S_{1,T}(\xi,\hat{\theta}_T(\tau))$ and $S_{2,T}(\xi,\hat{\theta}_T(\tau))$, say $S^*_{1,T}(\xi,\hat{\theta}^*_T(\tau))$ and $S^*_{2,T}(\xi,\hat{\theta}^*_T(\tau))$ respectively, are then defined to be
\begin{align} 
S^*_{1,T}(\xi,\hat{\theta}^*_T(\tau))&:= T^{-\frac{1}{2}}\sum_{t=1}^{T} \Big [\mathbbm{1}(y^*_t-Y^*_{t-1}\hat{\theta}^*_T(\tau)\leq 0)-\tau]exp(\xi'\Phi(\hat{I}^*_{t-1}))\\ \notag
& \quad \quad \quad \quad \quad \quad \quad \quad  -\mathbbm{1}(y_t-Y_{t-1}\hat{\theta}_T(\tau)\leq 0)-\tau]exp(\xi'\Phi(\hat{I}_{t-1}))     \Big ]\\  
S^*_{2,T}(\xi,\hat{\theta}^*_T(\tau))&:= T^{-\frac{1}{2}}\sum_{t=1}^{T} [\mathbbm{1}(y^*_t-Y^*_{t-1}\hat{\theta}^*_{1,T}(\tau)-\hat{F}^*_{t-1}\hat{\theta}^*_{2,T}(\tau)\leq 0)-\tau]exp(\xi'\Phi(\hat{I}^*_{t-1}))\\ \notag
& \quad \quad \quad \quad \quad \quad \quad \quad - \mathbbm{1}(y_t-Y_{t-1}\hat{\theta}_{1,T}(\tau)-\hat{F}_{t-1}\hat{\theta}_{2,T}(\tau)\leq 0)-\tau]exp(\xi'\Phi(\hat{I}_{t-1})) \Big].  
\end{align}\\

Consequently, one could then obtain the corresponding bootstrap functionals
\begin{itemize}
\item $CvM^*_{j,T}:=\int_{\mathcal{T}}\int_{\mathcal{Y}}|S^*_{j,T}(\xi,\hat{\theta}^*_T(\tau))|^2d\Phi_1(\xi)d\Phi_2(\tau)$
\item $KS^*_{j,T}:=sup_{\tau \in \mathcal{T}}\int_{\mathcal{Y}} |S^*_{j,T}(\xi,\hat{\theta}^*_T(\tau))|^2d\Phi_1(\xi)$.\\
\end{itemize}

For any bootstrap replication we compute the bootstrap functional $CvM^*_{j,T}$($KS^*_{j,T}$). Performing then B bootstrap replications, with B large, we compute the quantiles of the empirical distribution of the B bootstrap statistics.  The null hypothesis $H_{0,j}$ is rejected if $CvM_{j,T}$($KS_{j,T}$) based on the original sample is greater than the $(1-\alpha)^{th}$ percentile of the corresponding bootstrap distribution, where $\alpha$ is the level of significance. This is because $CvM_{j,T}$($KS_{j,T}$) has the same limiting distributions as its corresponding bootstrapped statistics, which ensures an asymptotic size equal to $\alpha$, while under the alternative $CvM_{j,T}$($KS_{j,T}$) diverges to infinity while the corresponding bootstrap statistics maintain its well defined distribution, ensuring asymptotic power. 

\vspace{1cm}

\section{Finite Sample Performance}
We consider two simulation cases to asses the finite sample performance of the proposed test statistics. The data generating process follows the following autoregressive process:\\
\begin{align*}
y_t=\rho_0(\upsilon_t)+\rho_1 (\upsilon_t)y_{t-1}+ \cdots +\rho_p (\upsilon_t)y_{t-p}+\beta_1F_{1,t-1}+\cdot + \beta_k(\upsilon_t)F_{k,t-1}
\end{align*}
where $\upsilon_t$ are independent Uniform $(0,1) $random variables . Motivated by our subsequent empirical application, the lag is taken as $p=1$ and the varying coefficients are defined as follows:\\
\begin{align*}
&\text{Case 1:} \rho_0 (\upsilon)=10+\Phi^{-1}(\upsilon),\rho_1(\upsilon)=0.5\\
&\text{Case 2:}\rho_0 (\upsilon)=10+\Phi^{-1}(\upsilon),\rho_1(\upsilon)=0.5, \beta_1(\epsilon)=0.8\\
\end{align*}
where the intercept coefficient $\rho_0(\upsilon)$ is from a normal distribution while the slope coefficients are constant. \\

As it is evident under Case 1 the null hypothesis $H_{0,1} $ is satisfied while under Case 2 the null hypothesis $H_{0,1}$  is violated but $H_{0,2}$ is satisfied. 
We consider multiple sample sizes, including $T=100$, $T=300$, $T=500$ and $T=1000$ , carry out $1000$ Monte Carlo Simulations, and for each of them perform $300$ bootstrap replications. In all the replications the nominal probability of rejecting a correct null hypothesis is 0.05. The conclusions with other nominal values are similar.\\

Following  the relevant literature, we choose the exponential function instead of the indicator function as it has been confirmed that exponential-based tests have higher performance than indicator-based tests. Furthermore, following the earlier relevant consistent test literature (see, e.g. \citet{bierens_consistent_1982}, \citet{bierens_consistent_1990}), we choose $\Phi$ as the $arctan$ function. In the experiment we consider a grid of $\mathcal{T}$ in $m=17$ equidistributed points from $\omega=0.1$ to $1- \omega=0.9$. Denote by $\mathcal{T}_m=\lbrace \tau_q \rbrace_{q=1}^m$ the point in the grid, with $w=\tau_1<\cdots<\tau_m=1-\omega$.  Also, as already mentioned we wish to weigh more heavily the more recent lags. Let $\Psi_{\tau}(\epsilon_t)=\mathbbm{1}(y_t-\rho_0(\tau)-\rho_1(\tau)y_{t-1}\leq 0)-\tau$ and let $exp(\xi'_Z \Phi(Z_t))=exp(\sum_{j=1}^d \xi'_j \Phi(Z_{t-j}))$, where $d=d(t)=\min \lbrace t-1,c \rbrace$ with $c<\infty$ and $\Xi={\xi_j: a_j\leq \xi_j \leq  \gamma_j, j=1,2; \vert a_j \vert, \vert \gamma_j \vert \leq \Gamma j^{-\kappa}, \kappa \geq 2}$.  It is immediate to see therefore that the weight attached to past observations decreases over time. We define $\Gamma$ over a grid, $\Gamma \in [0,3]$ and evaluate $g=30$ equidistributed points along the grid. Therefore the $CvM$ and $KS$ statistics are computed as:\\
\begin{align}
KS_T=sup_{\tau \in \mathcal{T}}  sup_{g \in \Gamma} \frac{1}{\sqrt{T}}\sum_{t=1}^T \Psi_{q}(\epsilon_t)exp\big(\sum_{j=0}^{d} \xi_j'(tan^{-1}(Y_{t-1})+tan^{-1}(\hat{F}_{t-1})\big) 
\label{KS Test Statistic}
\end{align}\\
\begin{align}
CvM_T=\frac{1}{mg^2\sqrt{T}}\sum_{q=1}^m \sum_{b=1}^g \sum_{t=1}^T \Psi_{q}(\epsilon_t)exp\big(\sum_{j=0}^{d} \xi_j'(tan^{-1}(Y_{t-1})+tan^{-1}(\hat{F}_{t-1})\big) 
\label{CvM Test Statistic}
\end{align}\\
The theory allows for $m \rightarrow \infty$ as $T \rightarrow \infty$  and the $\lbrace \tau_q \rbrace_{q=1}^m$  generated independently from a distribution on $\mathcal{T}$, but for simplicity in the computations we assume $m$ fixed and $\lbrace \tau_q \rbrace_{q=1}^m$ deterministic throughout the remainder of this paper.\\

\begin{table}[H]
\label{table:Monte Carlo Simulations}
\caption{Empirical size and power (rejection frequencies)}
\center
\begin{tabular}{lcllllll}
\hline
\hline
                                             & T      &  & \multicolumn{2}{c}{Case 1} &  & \multicolumn{2}{c}{Case 2} \\
                                             \hline
                                             &        &  & $KS_T$      & $CvM_T$      &  & $KS_T$       & $CvM_T$        \\  \cline{4-5} \cline{7-8}
                                             
\multirow{4}{*}{$\rho_1=0.5,\beta_1=0.8$}                     & 100  &  & 0.000      & 0.034      &  & 0.056       & 0.074         \\
                                             & 300  &  & 0.000      & 0.032       &  & 0.438       & 0.968        \\
                                             & 500  &  & 0.002      & 0.028       &  & 0.706      & 1.000 \\
                                             & 1000 &  & 0.006     & 0.036     &  &  0.950     &  1.000        \\
\hline  
\hline   
\end{tabular}
\end{table}
\bigskip

In Table 1 we report the rejection frequencies of the test based on the two continuous functionals, the \textit{Kolmogorov-Smirnov} and the \textit{Cramer-von-Mises}, for the model under the two data generating processes. The empirical size, though smaller than the nominal $5\%$  is satisfactory for the $CvM$ functional even with a low number of observations, in contrast with the $KS$ functional which remains severely undersized. With regards to the empirical power, we observe that the $CvM$  has the highest rejection frequencies, however both functionals capture the misspecification under the DGP of Case 2 relatively well, when the null hypothesis is that of a standard quantile autoregressive model with no additional factors (i.e. $H_{0,1}$).  This is in line with previous empirical results which have shown that the $CvM$ functional outperforms the $KS$ in terms of power \citep{escanciano_specification_2010}. Furthermore, these rejection rates increase with both the sample size and the strength of the alternative hypothesis imposed (i.e. with higher values of $\beta_1$).\\

The limited simulation study taken suggests that even with relatively small sample sizes (which might be a common scenario in empirical applications) the test exhibits fairly good size accuracy and power performance, particularly when using the \textit{CvM} functional.  As a result, in the subsequent empirical application we will be basing our analysis only on the \textit{Cramer-von-Mises} functional\footnote{It is worth nothing, that one could also employ the \textit{Kupiec} functional as well.}. \\ 

\vspace{1cm}

\section{Application To The GDP Growth And Inflation Distributions}
GDP growth and inflation are two of the most important variables due to their dominant role in many macroeconomic models (\citealp{levin_is_2002,angeloni_new_2006}). In the empirical literature, the majority of the work has examined how univariate regression can be improved by augmenting such models with factors as a way to summarise large amounts of information in estimates and forecasts of the conditional mean of these variables (\citealp{stock_forecasting_2002, bernanke_measuring_2005}). Nevertheless, such point forecasts ignore the risks around the central forecast. For example, in the case of growth a central forecast may paint an overly optimistic picture of the state of the economy, which is why the policymakers' focus on downside risk has increased in recent years \citep{adrian_vulnerable_2019}. Similarly, as argued by, inter alia, \citet{henry_is_2004}, the dynamic behaviour of inflation in particular has a number of economic implications. \\

\enquote{Fan charts} have been extensively used by the Bank of England in its Inflation Reports, to describe its best provision of future inflation to the general public since 1997, however, more recently, a number of inflation-targeting central banks have started to publish both GDP growth and inflation distributions. Examining the dynamics and higher moments of inflation and growth is of extreme importance and the ability to test whether these conditional quantile models should be augmented with latent factors representing a larger information set (e.g. financial conditions)  is necessary. 

\subsection{Data}
In this section, in an illustrative empirical analysis, we aim to examine whether the quantiles of the GDP growth rate and the CPI inflation rate are best modelled as univariate regressions or factor augmented models. We consider for the estimation of factors, data series containing data on inflation, real activity and indicators of money and key asset prices for the United Kingdom. We will employ the specification test using a sample which includes quarterly data from 177 macroeconomic variables, including the annual inflation and GDP growth rates, spanning from the second quarter of 1991 to the second quarter of 2018, with a total of $T=109$ observations. According to the criteria outlined in \citet{bai_determining_2002} the number of  common latent factors present in the data is one and for both of the dependent variables in consideration, the autoregressive model of order 1 has been deemed appropriate by the Schwarz criterion outlined in \citet{machado_robust_1993}.\\

\subsection{Specification Test}
We entertain for the GDP growth rate and the CPI inflation rate, both a Linear QAR model of order 1, as in (\ref{Linear QAR}), and a factor augmented QAR of order 1, with parameters estimated by quantile autoregression. In order to test $H_{0,1}: \quad E[\mathbbm{1}(y_t-\theta_0(\tau)-y_{t-1}\theta_1(\tau)) \leq 0)-\tau|I_{t-1}]=0$, a.s. versus $H_{A,1}: \quad Pr\{E[\mathbbm{1}(y_t-\theta_0(\tau)-Y_{t-1}\theta_1(\tau)) \leq 0)-\tau|I_{t-1}]=0\}$, we construct the $S_{1,T}$ statistics as defined in (\ref{Emprirical Process}), setting $I_{t-1}=[Y_{t-j},F_{1,t-j}]'$. Similarly in order to test $H_{0,2}: \quad E[\mathbbm{1}(y_t-\theta_0(\tau)-Y_{t-1}\theta_1(\tau)-F_{t-1}\theta_2(\tau)) -\leq 0)-\tau|I_{t-1}]=0$, a.s. versus $H_{A,1}: \quad Pr\{E[\mathbbm{1}(y_t-\theta_0(\tau)-y_{t-1}\theta_1(\tau)-f_{t-1}\theta_2(\tau)) \leq 0)-\tau|I_{t-1}]=0\}$, we construct the $S_{2,T}$ statistic. We use the exponential function, and set $\Phi$ as the inverse tangent function, while setting $\xi_j=\Gamma(j+1)^{-2}$, where $\Gamma$ is defined over a fine grid, $\Gamma=$ ( $\Gamma_1 \atop \Gamma_2$)$\in [0,3] \times [0,3]$. Similarly with the simulation setup we take $m$ equidistributed points $\lbrace \tau_g \rbrace_{g=1}^m$ from $\tau_1=0.1$ to $\tau_m=0.9$, but perform the test for multiple choices of $m$. In Table~\ref{table:Empirical Results} we report the p-values for the $CvM$ statistic obtained with the iid bootstrap with the number of bootstrap replications set to 500.\\

\vspace{0.5cm}

\begin{table}[H]
\center
\begin{tabular}{ccccccccc}
\hline
\hline
                                                                                & \textbf{m} & \textbf{QAR} & \textbf{FA-QAR} & \textbf{} &                                                                                & \textbf{m} & \textbf{QAR} \\
                                                                                \hline
\multirow{3}{*}{\textbf{\begin{tabular}[c]{@{}c@{}}Dependent Variable\\ GDP growth\end{tabular}}} & 5          & 0.006        & 0.330         &           & \multirow{3}{*}{\textbf{\begin{tabular}[c]{@{}c@{}}Dependent Variable\\ CPI Inflation\end{tabular}}} & 5          & 0.388     \\
                                                                                & 9          & 0.004        & 0.380 &           &                                                                                & 9          & 0.314   \\
                                                                                & 17         & 0.002        &  0.438 &           &                                                                                & 17         & 0.342  \\
                                                                                \hline
                                                                                \hline        
\end{tabular}
\caption{Bootstrapping p-values for the $CvM_{1,T}$ and $CvM_{2,T}$ functionals}
\label{table:Empirical Results}
\end{table}

\vspace{0.5cm}

From the results in Table~\ref{table:Empirical Results}, we can conclude that the $QAR(1)$ model of GDP growth is misspecified at $1\%$ nominal level, conditional on an information set that includes growth lags and the estimated factor, for all the number of quantiles estimated . We see that the omnibus test based on the $CvM_{1,T}$ strongly rejects this model and the rejection appears stronger when the grid over $\mathcal{T}$ becomes finer, a result that is consistent with the fact that the power of the test improves as $m \rightarrow \infty$. Our expectations that the factor augmented model should be an omitted variable in the QAR model is verified with the corresponding p-values, where we see that the misspecification is eliminated as we fail to reject the null hypothesis $H_{0,2}$ (the FA-QAR columns), which indicates that the factor-augmented QAR is correctly specified. On the other hand, we see that the $QAR(1)$ model of CPI Inflation is correctly specified and in fact the estimated latent factor is not an omitted variable and does not carry relevant information. This result is consistent with the literature supporting that past inflation is the most important determinant and thus best predictor for future inflation. \\

Based on these results we clearly see that estimated latent factors may often carry information that is relevant for the estimation of variables of interest and having a test which enables us to test if they are an omitted variable, is of critical importance in empirical analysis. We further complement the testing procedure by estimating multiple quantiles of growth and inflation, employing the Factor Augmented Quantile Autoregression and the Quantile Autoregression respectively, in an attempt to trace out their distributions. In addition we examine the impact of the independent variables of our regression on the dependent variable.\\ 

\begin{figure}[H]
\center
\includegraphics[scale=0.085]{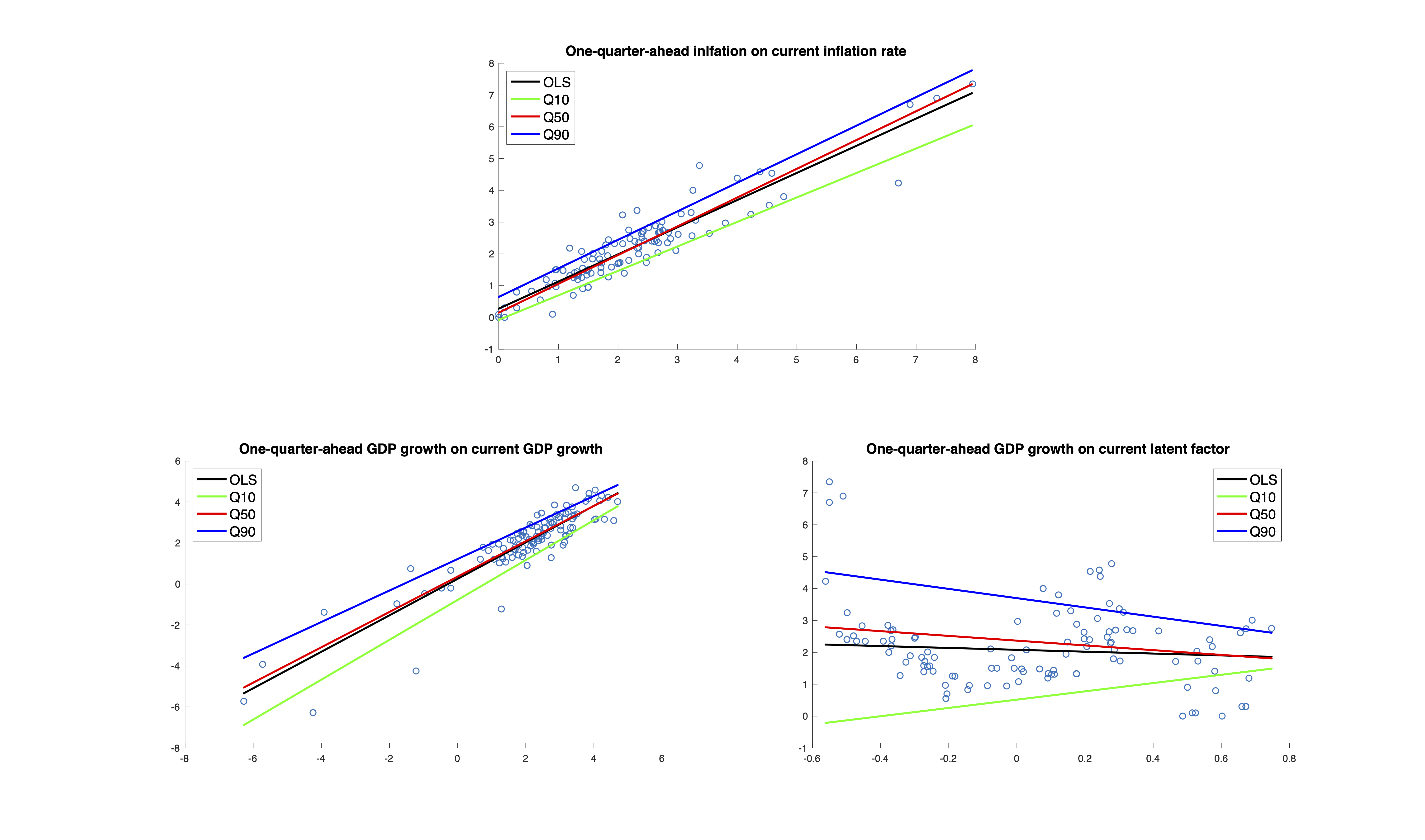}
\caption{Univariate Quantile Regressions}
\label{figure: Univariate Regressions}
\end{figure}

Figure~\ref{figure: Univariate Regressions} shows the scatter plots of one-quarter-ahead CPI inflation against the current inflation rate and one-quarter-ahead GDP growth against the current growth rate and the current latent factor, along with the corresponding univariate regressions for the tenth, fiftieth and ninetieth quantile and the least squares regression line. The slopes for the CPI inflation lag appear relatively stable  across the quantiles and so do the slopes of the GDP growth lag. Interestingly, the slopes of the latent factor differ significantly across the quantiles, implying that its impact is distinctively different across the spectrum of the growth distribution. Furthermore we see that for all the variables concerned the mean and median regression appear identical which one might consider as an indication of symmetry in the distribution of the dependent variables.

\begin{figure}[H]
\center
\captionsetup{justification=centering}
\includegraphics[scale=0.085]{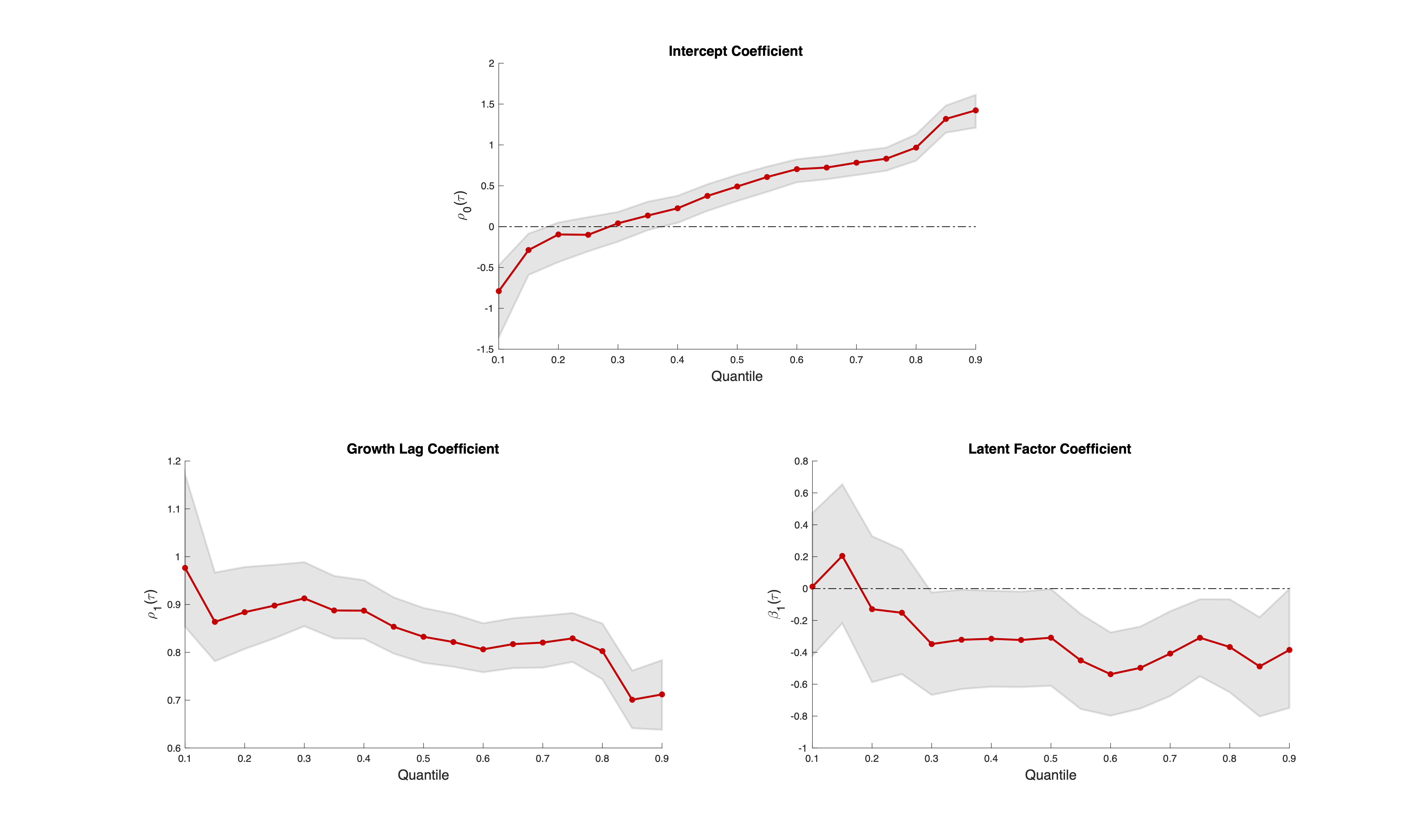}
\caption{Coefficients for the FA-QAR(1,1) model of GDP growth with $90\%$ confidence bands}
\label{figure: Coefficient Analysis GDP}
\end{figure}

\vspace{0.5cm}

\begin{figure}[H]
\center
\captionsetup{justification=centering}
\includegraphics[scale=0.085]{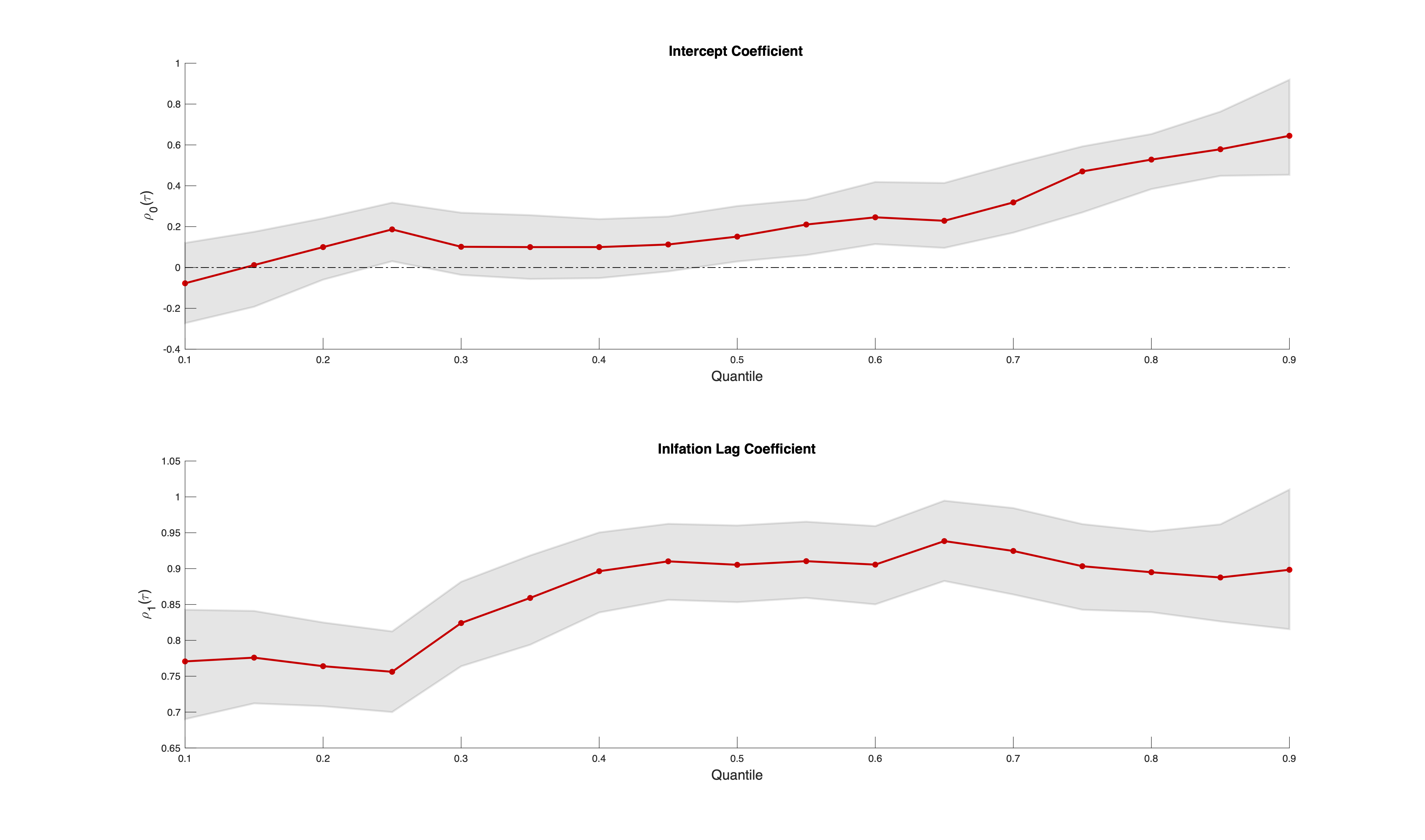}
\caption{Coefficients for the QAR(1) model of CPI inflation with $90\%$ confidence bands}
\label{figure: Coefficient Analysis CPI}
\end{figure}

Indeed, Figure~\ref{figure: Coefficient Analysis GDP} reveals that  the autoregressive coefficient of GDP growth changes slightly across the evaluated quantiles and appears to  become weaker across the upper tail of the distribution. This is not the case however for the latent factor coefficient. The current latent factor seems to be statistically not significant in the lower tails of the distribution, but its impact increases on the middle and towards upper tail of the distribution, as the coefficient becomes more negative. This implies that the latent factor is not uniformly informative for predicting tail outcomes. For lower quantiles of the GDP growth the lag coefficient is the sole determinant, however, as we start to consider the upper tail of the distribution, the latent factor seems to provide significant information. In the case of inflation, the autoregressive coefficient is statistically different from zero across all the estimated quantiles. The coefficient has a bigger impact on higher quantiles and in the extreme upper tail approaches a unit root process (see Figure~\ref{figure: Coefficient Analysis CPI}).\\

\vspace{0.5cm}

\begin{figure}[H]
\center
\captionsetup{justification=centering}
\includegraphics[scale=0.085]{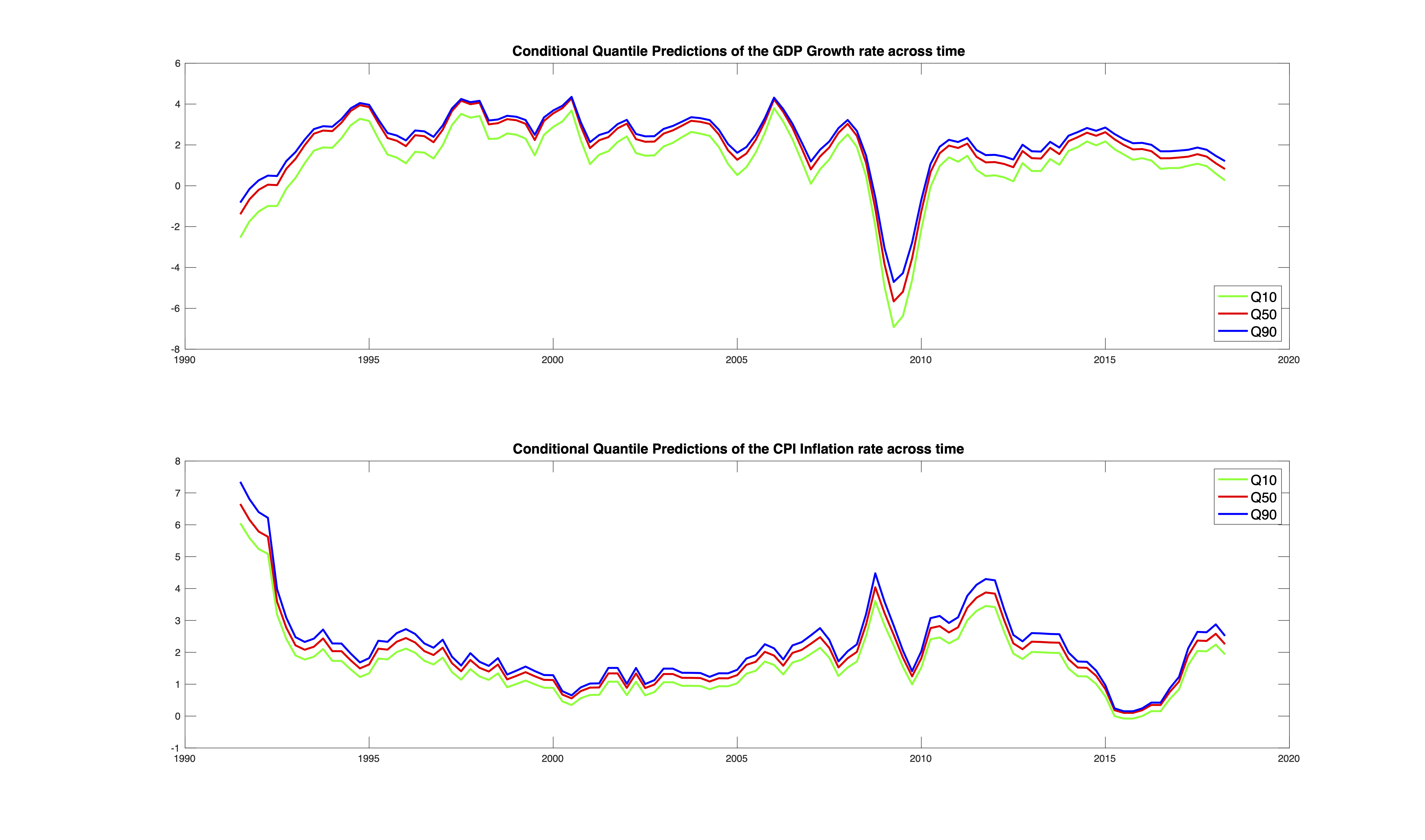}
\caption{Estimated Conditional Quantiles}
\label{figure: Conditional Quantile Predictions}
\end{figure}

\vspace{0.5cm}

Furthermore, in figure~\ref{figure: Conditional Quantile Predictions} we have traced out the tenth, fiftieth and ninetieth  conditional quantiles across the sample time span. The figure demonstrates three important empirical facts. On the one hand, if we focus on the periods before and around the 2008 economic crisis we can see an asymmetry in the distribution of GDP growth as the difference between the ninetieth quantile and the median is significantly smaller than the difference between the median and tenth quantile. This asymmetry however seems to dissipate following the recent financial crisis up until the end of the sample. Furthermore, the distribution of GDP growth has remained fairly stable across the sample, with the exception of the 2008 recession where GDP had experienced a significant decline. The distribution of CPI Inflation on the other hand has experienced smaller fluctuations throughout the sample and demonstrates a more symmetric behaviour. It is noticeable however, that the inflation distribution had narrowed significantly around 2015 when  the interest rate was close to the zero lower bound.\\ 

\begin{figure}[H]
\center
\includegraphics[scale=0.085]{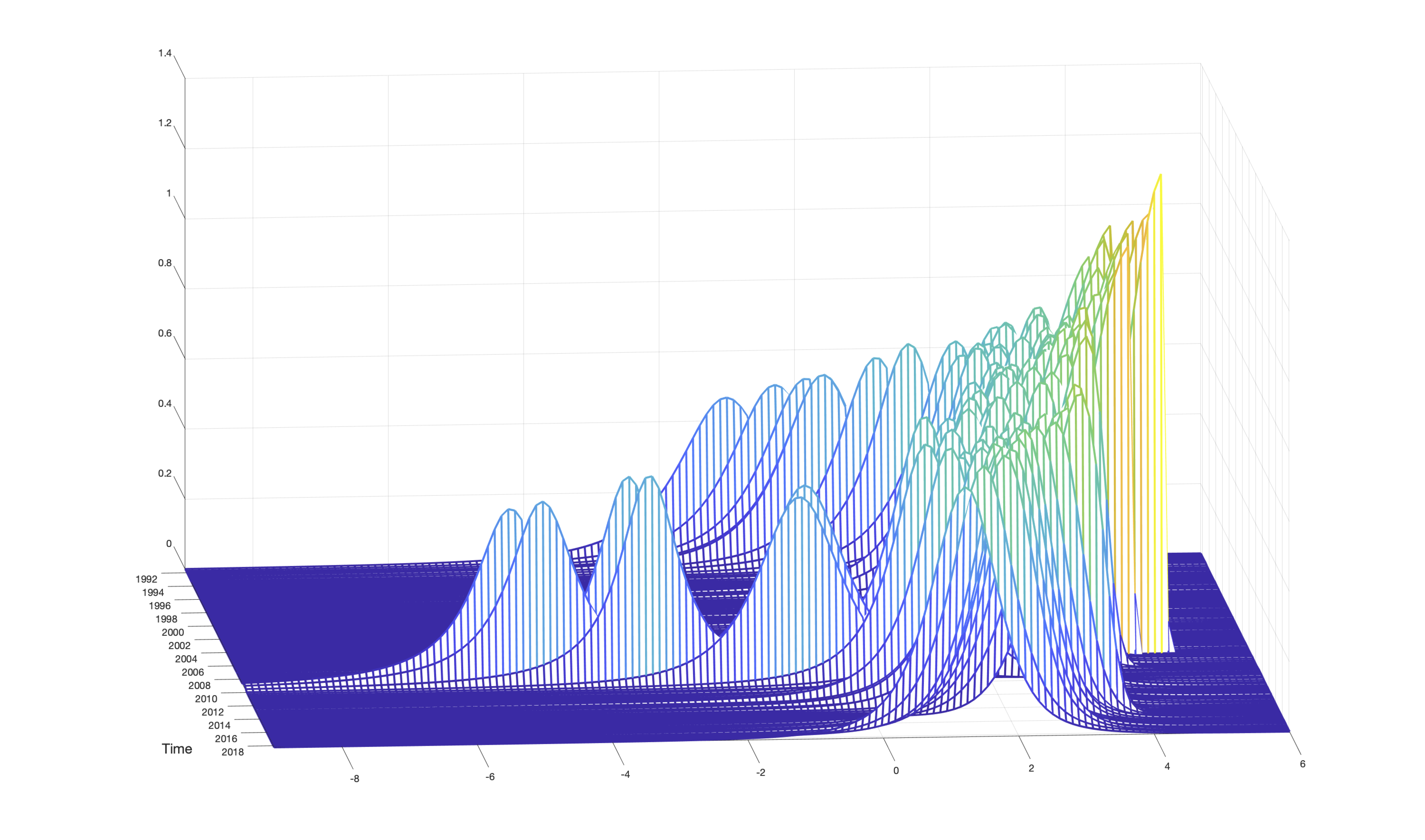}
\caption{Distribution of GDP growth over time}
\label{figure: Conditional Density GDP}
\end{figure}

\begin{figure}[H]
\center
\includegraphics[scale=0.085]{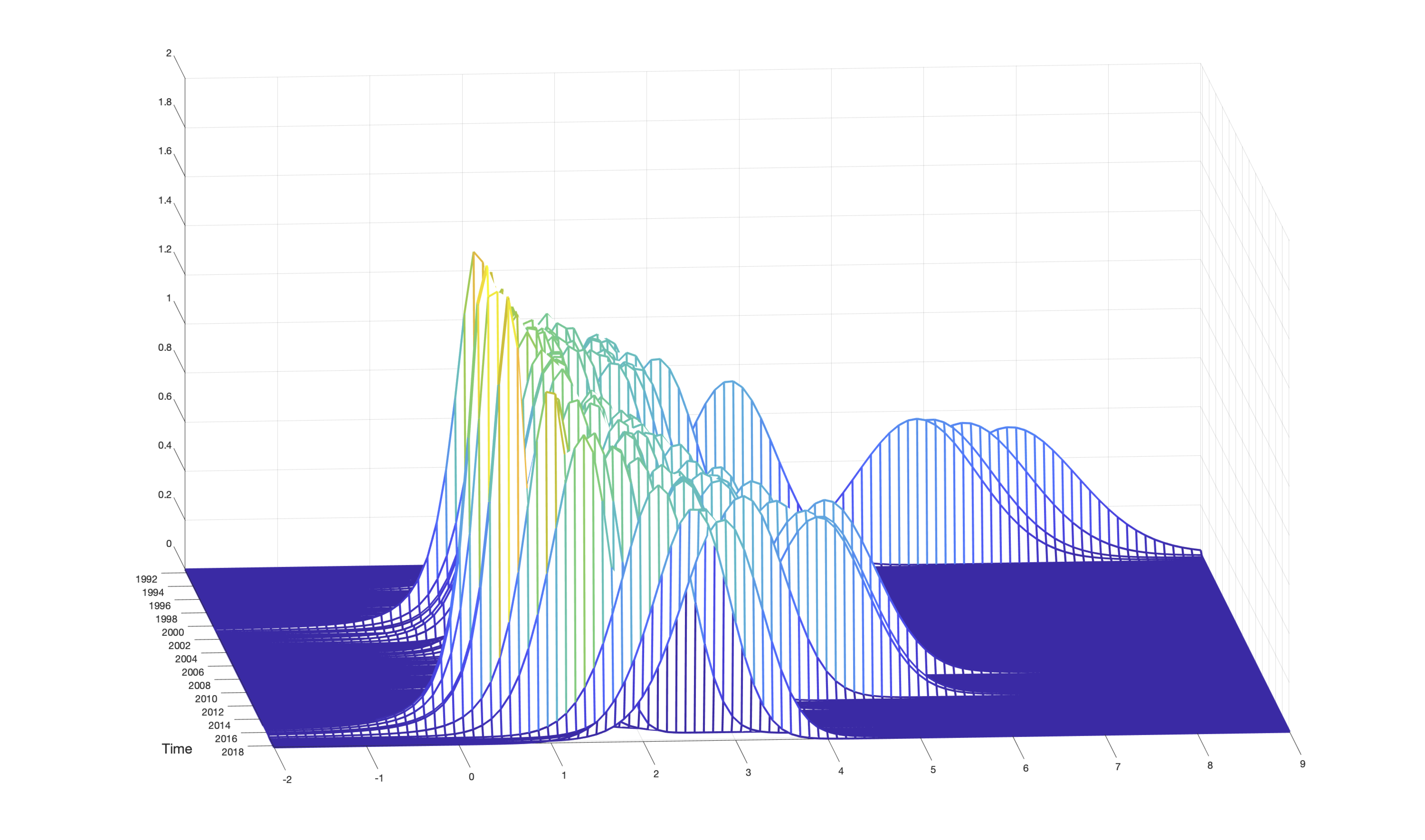}
\caption{Distribution of CPI inflation over time}
\label{figure: Conditional Density CPI}
\end{figure}

Based on these indications, we finally proceed to smooth the estimated quantile distributions each quarter by interpolating between the estimated quantiles using the skewed $t-distribution$ by \citet{azzalini_distributions_2003}, characterised by four parameters that pin down the location, $\mu$, scale,$\sigma$, fatness, $\nu$ and shape, $\alpha$. We fit the skewed $t-distribution$ in order to smooth the quantile function and recover a probability density function:
\begin{align}
f(y;\mu,\sigma,\alpha,\nu)=\frac{2}{\sigma} t\Big(\frac{y-\mu}{\sigma};\nu\Big) T\Big(\alpha \frac{y-\mu}{\sigma}\sqrt{\frac{\nu+1}{\nu+(\frac{y-\mu}{\sigma})^2}};\nu+1\Big)
\end{align}\\
where, $t(\cdot)$ and $T(\cdot)$ respectively denote the PDF and CDF of the student $t-$distribution. For each quarter we choose the four parameters $\lbrace \mu, \sigma,\nu,\alpha \rbrace$ of the skewed $t-distribution$ $f$ to minimise the squared distance between our estimated quantile function $m_j(I_{t-1}, \hat{\theta}(\tau))$ and the quantile function of the skewed $t-distribution$ $F^{-1}(\tau;\mu_t, \sigma_t,\nu_t,\alpha_t)$ to match the $5,25,75,95$ percent quantiles.This approach, first presented in \citet{adrian_vulnerable_2019}, is computationally less burdensome while also making fewer parametric assumptions compared to alternative ways of estimating conditional predictive distributions \citep{hamilton_new_1989, smith_asymmetric_2016} and it allows us to obtain an estimated conditional distribution of the GDP growth and CPI inflation, plotted in Figures~\ref{figure: Conditional Density GDP} $\&$ ~\ref{figure: Conditional Density CPI} .  \\

As it is evident in Figure~\ref{figure: Conditional Density GDP} the entire distribution and not just the central tendency of the GDP growth has evolved over time. For example the 2008 recession was associated with a rather symmetric distribution, albeit one with a significantly lower mean, while the antecedent expansion was associated with left skewed distributions. Furthermore the right tail of the distribution appears to be more stable than the median and lower tails which exhibit a stronger time variation.  The distinctive behaviour in the tails of the conditional distribution indicates that when it comes to growth downside risk varies much more strongly than upside risk over time, a fact that has also been found to be true for the US by \citep{adrian_vulnerable_2019}. The distribution of inflation on the other hand, with the exception of some outliers at the beginning of the sample, has remained fairly more stable over time. The fluctuations in the central tendency have been significantly smaller than the fluctuations in the tails of the distribution as well as the kyrtosis. Furthermore, we see that the left tail of the distribution has remained fairly stable, a fact associated with the zero lower bound of nominal interest rates, while the median and the right tail in particular exhibit stronger time series variation. We see therefore, that upside risk for inflation varies much more strongly and was a key ingredient during the 2008 financial crisis and for the following years, where as we see the right tail of the distribution had significantly shifted to the right. This in fact supports the thought that, though inflation had on average remain unaltered during the crisis, the inflation distribution was still affected by the volatility in the macro-economy in a multitude of ways, in terms of location, scale, fatness and shape. The significant variation in the distributions of both variables of interest demonstrate the importance of examining higher moments and dynamics rather than simple point forecasts which ignore the risks surrounding that  central forecast.\\

\vspace{0.5cm}

\vspace{0.5cm}

\vspace{1cm}

\section{Conclusion}
Econometric modelling often requires the specification of conditional quantile models for a range of quantiles of the conditional distribution. However, such models may often rely on unobservable variables that require estimation prior to modelling. For the evaluation therefore of quantile autoregressive models, we propose a test for the joint hypothesis of correct dynamic specification and no omitted latent variable with \enquote{valid} asymptotic critical values obtained via a bootstrap procedure based on the entire history of available information. We have demonstrated in a simulation study the consistency of the proposed test and its high power. An illustrative empirical implementation of the test suggests that the modelling of the conditional quantiles of UK GDP growth can be improved with the inclusion of estimated latent factors characterising the economy, while the CPI inflation rate does not require such additional information.  Furthermore an empirical analysis of the conditional dependence structure of both variables demonstrate that downside risk for growth exhibits higher fluctuations, while in the case of inflation upside risk becomes more relevant.\\

\newpage
\section{Appendix A}
\subsection{Assumptions on Latent Factors}
\textbf{Assumption A: Factors}\\
$E\Vert F_t \Vert^4\leq M<\infty$ and $T^{-1}\sum_{t=1}^T F_t F_t' \xrightarrow{\text{p}}\Sigma_F$ for some $k \times k$ positive definite matrix $\Sigma_F$.\\

\textbf{Assumption B: Factor loadings}\\
$E\Vert \lambda_i \Vert \leq \bar{\lambda} < \infty$ and $\Vert \frac{\Lambda\Lambda'}{N}-\Sigma_{\Lambda} \Vert  \rightarrow 0$ for some $k \times k$ positive definite matrix $\Sigma_{\Lambda}$.\\

\textbf{Assumption C: Time and cross-section dependence and heteroscedasticity}\\
There exists a positive constant $M < \infty$ such that for all $N$ and $T$,\\

\quad \quad 1. $E(e_{it})=0, E\vert e_{it}\vert^8 <M$;\\

\quad \quad 2. $E(e'_se_t/N)=E(N^{-1}\sum_{i=1}^N e_{is}e_{it})=\gamma_N(s,t), \vert \gamma_N(s,s) \vert \leq M$ for all s, and\\

\quad \quad \quad \quad \quad \quad \quad \quad $T^{-1}\sum_{s=1}^T\sum_{t=1}^T \vert \gamma_N(s,t) \vert \leq M$;\\

\quad \quad 3. $E(e_{it}e_{jt})=c_{ij,t}$ with $\vert c_{ij,t} \vert \leq\vert c_{ij} \vert$ for some $c_{ij}$ and for all $t$. In addition,\\

\quad \quad \quad \quad \quad \quad \quad \quad $N^{-1}\sum_{i=1}^N\sum_{j=1}^N\vert c_{ij}\vert \leq M$; \\

\quad \quad 4. $E(e_{it}e_{js})=c_{ij,ts}$ and $(NT)^{-1}\sum_{i=1}^N\sum_{j=1}^N\sum_{t=1}^T\sum_{s=1}^T \vert c_{ij,ts} \vert \leq M$;\\

\quad \quad 5. For every $(t,s)$, $E \vert N^{-\frac{1}{2}}\sum_{i=1}^N[e_{is}e_{it}-E(e_{is}e_{it})] \vert^4 \leq M$.\\

\textbf{Assumption D: Weak dependence between factors and idiosyncratic errors}\\

\quad \quad \quad \quad \quad \quad \quad \quad $E\big(\frac{1}{N}\sum_{i=1}^N \Vert \frac{1}{\sqrt{T}} \sum_{t=1}^T F_te_{it} \Vert^2 \big) \leq M$.\\

\textbf{Assumption E: Weak dependence}\\
There exists $M < \infty$ such that for all $T$ and $N$, and for every $t \leq T$ and every $i \leq N$,\\

\quad \quad 1. $\sum_{s=1}^{T} \vert \gamma_N(s,t) \vert \leq M$\\

\quad \quad 2. $\sum_{k=1}^{N} \vert c_{ki} \vert \leq M$\\

\textbf{Assumption F: Moments and central limit theorem}\\
There exists $M < \infty$ such that for all $N$, $T$\\

\quad \quad 1. For each t,\\

\quad \quad \quad \quad \quad \quad \quad \quad $E \Vert \frac{1}{\sqrt{NT}} \sum_{s=1}^{T} \sum_{k=1}^{N} F_t[e_{ks}e_{kt}-E(e_{ks}e_{kt})] \Vert^2 \leq M$\\

\quad \quad 2. The $k \times k$ matrix satisfies\\

\quad \quad \quad \quad \quad \quad \quad \quad $E \Vert \frac{1}{\sqrt{NT}} \sum_{t=1}^{T} \sum_{k=1}^{N} F_t \lambda_k' e_{kt} \Vert^2 \leq M$\\

\quad \quad 3. For each $t$, as $N \rightarrow N$,\\

\quad \quad \quad \quad \quad \quad \quad \quad $\frac{1}{\sqrt{N}}\sum_{i=1}^{N} \lambda_i e_{it} \xrightarrow{\text{d}} N(0, \Gamma_t)$\\

\quad \quad where $\Gamma_t= \lim\limits_{N \rightarrow \infty} \frac{1}{N}\sum_{i=1}^{N}\sum_{j=1}^{N} \lambda_i \lambda_j E(e_{it}e_{jt})$\\

\quad \quad 4. For each $i$, as $T \rightarrow \infty$,\\

\quad \quad \quad \quad \quad \quad \quad \quad $\frac{1}{\sqrt{T}}\sum_{t=1}^{T} F_t e_{it} \xrightarrow{\text{d}} N(0, W_i)$,\\

\quad \quad where $W_i= \underset{T \to \infty}{\plim} \frac{1}{T}\sum_{s=1}^{T} \sum_{t=1}^{T} E[F_tF_s'e_{is}e_{it}]$.\\

Assumption A is more general than that of classical factor analysis, as it allows $F_t$ to be dynamic such that $A(L)F_t=v_t$. However, this dynamic does not enter into $X_{it}$ directly, so the relationship between $X_{it}$ and $F_t$ remains static. Assumption B ensures that each factor has a non-trivial contribution to the variance of $X_t$. Assumption C allows for limited time series and cross section dependence in the idiosyncratic components as well as heteroscedasticity in both dimensions. Assumption D does not require independence to hold and is implied by Assumptions A and C. Throughout this article the number of factors ($k$) is assumed fixed as $N$ and $T$ grow. Assumptions A-D are sufficient for consistently estimating the number of factors ($k$), as well as the factors themselves and their corresponding loadings. Assumption E is a stronger version of assumptions C2 and C3 but still reasonable. For example, under time and cross sectional independence assumptions E1 and E2 become equivalent since  $\frac{1}{N} \sum_{i=1}^N (e_{it})^2 \leq M$ and $E(e_{it})^2 \leq M$ which are both implied by assumption C1.  Similarly, Assumption F is not stringent because sums in F1 and F2 involve zero mean random variables. The last two assumptions are central limit theorems, which are satisfied by several mixing processes.\\

\vspace{1cm}

\subsection{Theorems $\&$ Proofs}
\small{
\textbf{Proof of Lemma 1.} As outlined in the main text, in practice the true factors $F$ in our information set are unobservable and have to be replaced by their estimated counterparts $\hat{F}$. Therefore,once $\hat{F}_{t-1}$ are replaced in $I_{t-1}$ resulting in $\hat{I}_{t-1}$, we demonstrate  how this factor estimation error affects the asymptotic properties of $S_{j,T}(\xi,\tau,\theta)$.  The proof will be dealing with $S_{2,T}(\xi,\tau,\theta)$, where factor estimation error is present both within the model and the information set (thus both in the indicator function and the weighting exponential function) hence $m_j(\hat{I}_{t-1},\hat{\theta}(\tau))=Y_{t-1}\hat{\theta}_{1,T}(\tau)-\hat{F}_{t-1}\hat{\theta}_{2,T}(\tau)$. It is easy to see that the results also hold for $S_{1,T}(\xi,\tau,\theta)$ where the factor estimation error is present only in the information set. \\

\vspace{0.5cm}

Our results will utilise the following identity, outlined in \citet{bai_determining_2002},
\begin{align}
\begin{split}
\hat{F}_t-H'F_t=V_{NT}^{-1}\big[\frac{1}{T}\sum_{s=1}^{T}\hat{F}_s\gamma_N(s,t)+\frac{1}{T}\sum_{s=1}^{T}\hat{F}_s\zeta_{st}+\frac{1}{T}\sum_{s=1}^{T}\hat{F}_s\eta_{st}+\frac{1}{T}\sum_{s=1}^{T}\hat{F}_s\xi_{st}\big]
\end{split}
\end{align}\begin{align*}
\text{where} \quad \quad \quad \quad \quad \quad \quad \quad \quad \gamma_N(s,t)&= E(\frac{1}{N}\sum_{s=1}^{N}e_{is}e_{it})= E\big(\frac{e'_se_t}{N}\big)\quad \quad \quad \quad \quad \quad \quad \quad \quad \quad \quad \quad\\
\zeta_{st}&=\frac{e'_se_t}{N}-\gamma_N(s,t)\\
\eta_{st}&=\frac{F'_s\Lambda'e_t}{N}\\
\xi_{st}&=\frac{F'_t\Lambda'e_s}{N}
\end{align*}\\

\vspace{0.5cm}

We will also utilise the following lemmas proved in \citet{stock_forecasting_1999} and \citet{bai_determining_2002} respectively:\\
\textbf{Lemma A.1:} Under Assumptions A-D, as $T$, $N \rightarrow \infty$,\\

($i$) $T^{-1} \hat{F}'(\frac{1}{TN}X'X)\hat{F}=V_{NT} \xrightarrow{\text{p}} V_{NT}$\\

($ii$) $\frac{\hat{F}'F}{T}(\frac{\hat{\Lambda}'\Lambda}{N})\frac{F'\hat{F}}{T}\xrightarrow{\text{p}} V$\\
where $V$ is the diagonal matrix consisting of the eigenvalues of $\Sigma_{\Lambda}\Sigma_F$.\\

\vspace{0.5cm}

\textbf{Lemma A.2:} Under assumptions A-D,\\

\quad \quad \quad \quad \quad \quad \quad \quad $\delta_{NT}^2\Big(\frac{1}{T}\sum_{t=1}^{T} \Vert \hat{F}_t-H'F_t \Vert^2  \Big)=O_p(1)$\\

Also note that because $V_{NT}$ converges to a positive definite matrix, by Lemma A.1, it follows that $\Vert V_{NT} \Vert = O_p(1)$. Furthermore, under assumptions A-B with $\hat{F}'F/T=I$ and Lemma A.1 , it is implied that $\Vert H \Vert = O_p(1)$.\\

Let hereinafter $\tilde{\epsilon}_t(\tau)=y_t-Y_{t-1}\hat{\theta}_{1,T}(\tau)-\hat{F}_{t-1}\hat{\theta}_{2,T}(\tau)$ be the residual when both factor estimation error and parameter estimation error is present. Also, let $\hat{\epsilon}_t(\tau)=y_t-Y_{t-1}\hat{\theta}_{1,T}(\tau)-F_{t-1} \hat{\theta}_{2,T}(\tau)$ be the residual with only parameter estimation error\footnote{Note that a similar residual is obtained under the QAR model and thus in the statistic $S_{1,T}(\xi,\hat{\theta}(\tau))$.}. Therefore, the quantile empirical process could be expressed in the following way:
\begin{align*}
\begin{split}
S_{2,T}(\xi,\hat{\theta}(\tau))&= T^{-\frac{1}{2}}\sum_{t=1}^{T} [\mathbbm{1}(y_t-Y_{t-1}\hat{\theta}_{1,T}(\tau)-\hat{F}_{t-1}\hat{\theta}_{2,T}(\tau)\leq 0)-\tau]exp(\xi'\Phi(\hat{I}_{t-1}))\\
&=T^{-\frac{1}{2}}\sum_{t=1}^{T} [\mathbbm{1}(\tilde{\epsilon}_t(\tau)\leq 0)-\tau]exp(\xi'\Phi(\hat{I}_{t-1}))\\
&=T^{-\frac{1}{2}}\sum_{t=1}^{T} [\mathbbm{1}(\hat{\epsilon}_t(\tau)\leq \hat{\epsilon}_t(\tau)-\tilde{\epsilon}_t(\tau))-\tau]exp(\xi'\Phi(\hat{I}_{t-1}))\\
\end{split}
\end{align*}\\

For simplicity, we are going to drop the dependence of $\epsilon$ and $e_t$ on the specified quantile $\tau$. By adding and subtracting $\mathbbm{1}(\hat{\epsilon}_t\leq 0)$ we obtain:\\
\begin{align*}
\begin{split}
S_{2,T}(\xi,\hat{\theta}(\tau))&=\underbrace{T^{-\frac{1}{2}}\sum_{t=1}^{T} [\mathbbm{1}(\hat{\epsilon}_t\leq 0)-\tau]exp(\xi'\Phi(\hat{I}_{t-1}))}_{\text{Term I}} +\underbrace{T^{-\frac{1}{2}}\sum_{t=1}^{T} [\mathbbm{1}(\hat{\epsilon}_t\leq \hat{\epsilon}_t-\tilde{\epsilon}_t)-\mathbbm{1}(\hat{\epsilon}_t\leq 0)]exp(\xi'\Phi(\hat{I}_{t-1}))}_{\text{Term II}}\\
\end{split}
\end{align*}\\

Focusing on \textbf{Term I\footnote{It is worth mentioning here that \textbf{Term I} is equivalent to the term that would be obtained under $S_{1,T}(\xi,\hat{\theta}(\tau))$, where estimated factors only appear in the conditioning set and thus the indicator function only includes parameter estimation error.  Solely proving that factor estimation error is negligible in \textbf{Term I} suffices to prove Lemma 1 for $S_{1,T}(\xi,\hat{\theta}(\tau))$.}}, using an intermediate value expansion around the true factors we obtain the following:
\begin{align*}
\begin{split}
I&= T^{-\frac{1}{2}}\sum_{t=1}^{T} [\mathbbm{1}(y_t-Y'_{t-1}\theta(\tau)\leq 0)-\tau]exp(\xi'\Phi(\hat{I}_{t-1}))\\
&= T^{-\frac{1}{2}}\sum_{t=1}^{T} [\mathbbm{1}(\epsilon_t(\tau)\leq 0)-\tau]exp(\xi_{Y}'\Phi({Y}_{t-1}))exp(\xi_{F}'\Phi(\hat{F}_{t-1}))
\\
&= T^{-\frac{1}{2}}\sum_{t=1}^{T}\Psi_{\tau}(\epsilon_t) exp(\xi_{Y}'\Phi({Y}_{t-1})) \large[exp(\xi_{F}'\Phi({HF}_{t-1})+\large{\bigtriangledown}_{F|\overline{HF}}(exp(\xi_{F
}'\Phi(\hat{F}_{t-1})))[\hat{F}_{t-1}-HF_{t-1}] \large]
\\
&= \underbrace{T^{-\frac{1}{2}}\sum_{t=1}^{T}\Psi_{\tau}(\epsilon_t) exp(\xi_{Y}'\Phi({Y}_{t-1}))exp(\xi_{F}'\Phi({HF}_{t-1}))}_{I_1} \quad +\\
&\quad  \quad \quad \quad \underbrace{T^{-\frac{1}{2}}\sum_{t=1}^{T}\Psi_{\tau}(\epsilon_t) exp(\xi_{Y}'\Phi({Y}_{t-1})){\bigtriangledown}_{F|\overline{HF}}(exp(\xi_{F}'\Phi(\hat{F}_{t-1})))*[\hat{F}_{t-1}-HF_{t-1}]}_{I_2}
\end{split}
\end{align*}\\

\textbf{Term I\textsubscript{1}} is equivalent to a test statistic where the information set includes the true latent factors and thus only has parameter estimation error present\footnote{This term is  equivalent to the one in \citet{escanciano_specification_2010}, with the exception that the imaginary unit is used instead of the Borel measurable function, $\Phi$. } and is therefore the test statistic with no factor estimation error. Focusing then on the \textbf{Term I\textsubscript{2}}, using the identity of the forecast error as that is shown in \citet{bai_determining_2002} we get:
\begin{align*}
\begin{split}
I_2&=T^{-\frac{1}{2}}\sum_{t=1}^{T} [\mathbbm{1}(\hat{\epsilon}_t\leq 0)-\tau] \bigtriangledown_{\hat{I}_{t-1}|\bar{I}_{t-1}}(exp(\xi'\Phi(\hat{I}_{t-1})))(\hat{F}_{t-1}-H'F_{t-1})\\
&=T^{-\frac{1}{2}}\sum_{t=1}^{T} [\mathbbm{1}(\hat{\epsilon}_t\leq 0)-\tau] \bigtriangledown_{\hat{F}_{t-1}|\overline{H'F}_{t-1}}(exp(\xi'\Phi(\hat{I}_{t-1})))\\
&\quad \quad \quad \quad \quad \quad \quad \quad*\Big\lbrace V_{NT}^{-1} \Big[T^{-1}\sum_{s=1}^T \hat{F}_s \gamma_N(s,t)+T^{-1}\sum_{s=1}^T \hat{F}_s \zeta_{st}+T^{-1}\sum_{s=1}^T \hat{F}_s \eta_{st} +T^{-1}\sum_{s=1}^T \hat{F}_s \xi_{st}  \Big]  \Big\rbrace\\
& \leq \sup_{t} [\bigtriangledown_{\hat{F}_{t-1}|\overline{H'F}_{t-1}}(exp(\xi'\Phi(\hat{I}_{t-1})))]  \underbrace{\Vert V_{NT}^{-1} \Vert}_{O_p(1)}T^{-\frac{3}{2}}\sum_{t=1}^{T}\sum_{s=1}^T [\mathbbm{1}(\hat{\epsilon}_t\leq 0)-\tau] \Big[\hat{F}_s \gamma_N(s,t)+\hat{F}_s \zeta_{st}+\hat{F}_s \eta_{st}+\hat{F}_s\xi_{st} \Big]
\end{split}
\end{align*}\\

Neglecting $V^{-1}_{NT}$ and also $\sup_{t} [\bigtriangledown_{\hat{F}_{t-1}|\overline{H'F}_{t-1}}(exp(b'\Phi(\hat{I}_{t-1})))] $, which is bounded by definition and accounting for the fact that $T^{-1}\sum_{s=1}^T \Vert \hat{F}_s \Vert^2=O_p(1)$, under assumption E1, the first term of \textbf{Term I\textsubscript{2}} is bounded by:
\begin{align*}
\begin{split}
T^{-\frac{3}{2}}\sum_{t=1}^{T}\sum_{s=1}^T \vert \mathbbm{1}(\hat{\epsilon}_t\leq 0)-\tau\vert \Vert \hat{F}_s \gamma_N(s,t)\Vert &\leq \underbrace{sup_t \vert \mathbbm{1}(\hat{\epsilon}_t\leq 0)-\tau\vert}_{\leq M \text{ by definition}} T^{-\frac{1}{2}}\sum_{t=1}^{T}\sum_{s=1}^T\Vert \hat{F}_s \gamma_N(s,t)\Vert\\
& \leq M*T^{-\frac{3}{2}} \Big(T^{-1}\sum_{s=1}^T \Vert \hat{F}_s \Vert^2 \Big)^{\frac{1}{2}} \Big(T^{-1}\sum_{s=1}^T \Big( \sum_{t=1}^T \vert \gamma_N(s,t) \vert\Big)^2 \Big)^{\frac{1}{2}} \\
&=O_p\Big(\frac{1}{\sqrt{T}} \Big)
\end{split}
\end{align*}\\

Similarly, under assumption C5 the second term is bounded by:
\begin{align*}
\begin{split}
T^{-\frac{3}{2}}\sum_{t=1}^{T}\sum_{s=1}^T \vert \mathbbm{1}(\hat{\epsilon}_t\leq 0)-\tau\vert \Vert \hat{F}_s \zeta_{st}\Vert &\leq \underbrace{sup_t \vert \mathbbm{1}(\hat{\epsilon}_t\leq 0)-\tau\vert}_{\leq M \text{ by definition}} T^{-\frac{3}{2}}\sum_{t=1}^{T}\sum_{s=1}^T\Vert \hat{F}_s \zeta_{st}\Vert\\
& \leq M*\sqrt{T} \Big(T^{-1}\sum_{s=1}^T \Vert \hat{F}_s \Vert^2 \Big)^{\frac{1}{2}} \Big(T^{-1}\sum_{s=1}^T \Big(T^{-1} \sum_{t=1}^T \vert \zeta_{st} \vert\Big)^2 \Big)^{\frac{1}{2}} \\
& \leq M*\frac{\sqrt{T}}{\sqrt{N}} \Big(\underbrace{T^{-1}\sum_{s=1}^T \Vert \hat{F}_s \Vert^2}_{O_p(1)} \Big)^{\frac{1}{2}} \Big(T^{-1}\sum_{s=1}^T \Big(\underbrace{ T^{-1}\sum_{t=1}^T \vert \frac{1}{\sqrt{N}}\sum_{i=1}^N e_{is}e_{it}-\gamma_N(s,t) \vert}_{O_p(1)}\Big)^2 \Big)^{\frac{1}{2}} \\
&=O_p\Big(\frac{\sqrt{T}}{\sqrt{N}} \Big)
\end{split}
\end{align*}\\

Furthermore, given that $T^{-1}\sum_{s=1}^T \Vert \hat{F}_s \Vert^2=O_p(1)$, under assumptions A and F3, we can obtain an upper bound for the third and fourth term term:
\begin{align*}
\begin{split}
T^{-\frac{3}{2}}\sum_{t=1}^{T}\sum_{s=1}^T \vert \mathbbm{1}(\hat{\epsilon}_t\leq 0)-\tau\vert \Vert \hat{F}_s \eta_{st}\Vert &\leq \underbrace{sup_t \vert \mathbbm{1}(\hat{\epsilon}_t\leq 0)-\tau\vert}_{\leq M \text{ by definition}} T^{-\frac{3}{2}}\sum_{t=1}^{T}\sum_{s=1}^T\Vert \hat{F}_s \eta_{st}\Vert\\
& \leq M*\sqrt{T} \Big(T^{-1}\sum_{s=1}^T \Vert \hat{F}_s \Vert^2 \Big)^{\frac{1}{2}} \Big(T^{-1}\sum_{s=1}^T \Big(T^{-1} \sum_{t=1}^T \Vert \frac{F_s' \Lambda' e_t}{N} \Vert\Big)^2 \Big)^{\frac{1}{2}} \\
& \leq M*\frac{\sqrt{T}}{\sqrt{N}} \Big(\underbrace{T^{-1}\sum_{s=1}^T \Vert \hat{F}_s \Vert^2}_{O_p(1)} \Big)^{\frac{1}{2}} \Big( \underbrace{(T^{-1}\sum_{s=1}^T \Vert F_s \Vert^2}_{O_p(1)}) * (\underbrace{ T^{-1}\sum_{t=1}^T \Vert \frac{\Lambda' e_t}{\sqrt{N}} \Vert}_{O_p(1)})^2\Big) \Big)^{\frac{1}{2}} \\
&=O_p\Big(\frac{\sqrt{T}}{\sqrt{N}} \Big)
\end{split}
\end{align*}\\

\begin{align*}
\begin{split}
T^{-\frac{3}{2}}\sum_{t=1}^{T}\sum_{s=1}^T \vert \mathbbm{1}(\hat{\epsilon}_t\leq 0)-\tau\vert \Vert \hat{F}_s \eta_{st}\Vert &\leq \underbrace{sup_t \vert \mathbbm{1}(\hat{\epsilon}_t\leq 0)-\tau\vert}_{\leq M \text{ by definition}} T^{-\frac{3}{2}}\sum_{t=1}^{T}\sum_{s=1}^T\Vert \hat{F}_s \eta_{st}\Vert\\
& \leq M*\sqrt{T} \Big(T^{-1}\sum_{s=1}^T \Vert \hat{F}_s \Vert^2 \Big)^{\frac{1}{2}} \Big(T^{-1}\sum_{s=1}^T \Big(T^{-1} \sum_{t=1}^T \Vert \frac{F_t' \Lambda' e_s}{N} \Vert\Big)^2 \Big)^{\frac{1}{2}} \\
& \leq M*\frac{\sqrt{T}}{\sqrt{N}} \Big(\underbrace{T^{-1}\sum_{s=1}^T \Vert \hat{F}_s \Vert^2}_{O_p(1)} \Big)^{\frac{1}{2}} \Big( \underbrace{(T^{-1}\sum_{s=1}^T \Vert \frac{\Lambda' e_s}{\sqrt{N}} \Vert^2}_{O_p(1)}) * (\underbrace{ T^{-1}\sum_{t=1}^T \Vert F_t \Vert}_{O_p(1)})^2\Big) \Big)^{\frac{1}{2}} \\
&=O_p\Big(\frac{\sqrt{T}}{\sqrt{N}} \Big)
\end{split}
\end{align*}\\
Overall therefore, given that $N \geqslant T$,
\begin{align*}
\textbf{Term I\textsubscript{2}}&=O_p\Big(\frac{1}{\sqrt{T}} \Big)+O_p\Big(\frac{\sqrt{T}}{\sqrt{N}} \Big) +O_p\Big(\frac{\sqrt{T}}{\sqrt{N}} \Big)+O_p\Big(\frac{\sqrt{T}}{\sqrt{N}}\Big)\\
&=O_p\Big(\frac{\sqrt{T}}{\min\lbrace\sqrt{N},T\rbrace}\Big)
\end{align*}\\

Moving on to \textbf{Term II:} By adding and subtracting $F(0)$ and $F(\hat{\epsilon}_t-\tilde{\epsilon})$, we obtain the following:
\begin{align*}
\begin{split}
II&=T^{-\frac{1}{2}}\sum_{t=1}^{T} [\mathbbm{1}(\hat{\epsilon}_t\leq \hat{\epsilon}_t-\tilde{\epsilon}_t)-\mathbbm{1}(\hat{\epsilon}_t\leq 0)]exp(\xi'\Phi(\hat{I}_{t-1}))\\
&= T^{-\frac{1}{2}}\sum_{t=1}^{T} \Big \lbrace [\mathbbm{1}(\hat{\epsilon}_t\leq \hat{\epsilon}_t-\tilde{\epsilon}_t)-F(\hat{\epsilon}_t-\tilde{\epsilon})] -[T^{-\frac{1}{2}}\sum_{t=1}^{T}\mathbbm{1}(\hat{\epsilon}_t\leq 0) -F(0)] \Big \rbrace exp(\xi'\Phi(\hat{I}_{t-1}))\\
&\quad \quad \quad \quad \quad \quad \quad \quad \quad \quad \quad \quad \quad \quad \quad \quad  \quad \quad \quad \quad \quad \quad \quad + T^{-\frac{1}{2}}\sum_{t=1}^{T} [F(\hat{\epsilon}_t-\tilde{\epsilon})-F(0)]exp(\xi'\Phi(\hat{I}_{t-1}))\\
& \leq \sup_t \Vert exp(\xi'\Phi(\hat{I}_{t-1})) \Vert \Big  \lbrace T^{-\frac{1}{2}}\sum_{t=1}^{T} [\mathbbm{1}(\hat{\epsilon}_t)\leq \hat{\epsilon}_t-\tilde{\epsilon}_t)-F(\hat{\epsilon}_t-\tilde{\epsilon})] -T^{-\frac{1}{2}}\sum_{t=1}^{T}[\mathbbm{1}(\hat{\epsilon}_t\leq 0) -F(0)] \Big \rbrace\\
&\quad \quad \quad \quad \quad \quad \quad \quad \quad \quad \quad \quad \quad \quad \quad \quad  \quad \quad \quad \quad \quad \quad \quad + T^{-\frac{1}{2}}\sum_{t=1}^{T} [F(\hat{\epsilon}_t-\tilde{\epsilon})-F(0)]exp(\xi'\Phi(\hat{I}_{t-1}))\\
&=T^{-\frac{1}{2}}\sum_{t=1}^{T} [F(\hat{\epsilon}_t-\tilde{\epsilon})-F(0)]exp(\xi'\Phi(\hat{I}_{t-1})+o_p(1)
\end{split}
\end{align*}\\
The last equality arises by stochastic equicontinuity and the bounded nature of the first component of the first term. \\

Similar to before, we can employ an intermediate value expansion on the remaining term and thus  obtain the following: 
\begin{align*}
\begin{split}
T&^{-\frac{1}{2}}\sum_{t=1}^{T} [F(\hat{\epsilon}_t-\tilde{\epsilon})-F_{\hat{\epsilon}}(0)]exp(\xi'\Phi(\hat{I}_{t-1}))\\
&=  T^{-\frac{1}{2}}\sum_{t=1}^{T} \Big[ F \Big((Y_t-Y'_{t-1}\hat{\theta}(\tau)-H'F'_{t-1}\hat{\beta}(\tau))- (Y_t-Y'_{t-1}\hat{\theta}(\tau)-\hat{F}'_{t-1}\hat{\beta}(\tau))\Big)-F(0)  \Big]exp(\xi'\Phi(\hat{I}_{t-1}))\\
&=  T^{-\frac{1}{2}}\sum_{t=1}^{T} \Big[ F \Big(\underbrace{(Y_t-Y'_{t-1}\hat{\theta}(\tau)-(H'F_{t-1})'\hat{\beta}(\tau))- (Y_t-Y'_{t-1}\hat{\theta}(\tau)-(H'F_{t-1})'\hat{\beta}(\tau))}_{=0} \Big) -F(0) \Big)\\
&\quad  +\bigtriangledown_{H'F}F \Big((Y_t-Y'_{t-1}\hat{\theta}(\tau)-(H'F'_{t-1})'\hat{\beta}(\tau))- (Y_t-Y'_{t-1}\hat{\theta}(\tau)-(\overline{H'F}_{t-1})'\hat{\beta}(\tau))\Big)*(\hat{F}_{t-1}-H'F_{t-1}) exp(\xi'\Phi(\hat{I}_{t-1}))\\
&= T^{-\frac{1}{2}}\sum_{t=1}^{T} \bigtriangledown_{H'F}F \Big(-((H'F_{t-1})'-(\overline{H'F}_{t-1})')\hat{\beta}(\tau)\Big)*(\hat{F}_{t-1}-H'F_{t-1}) exp(\xi'\Phi(\hat{I}_{t-1}))\\
&= \underbrace{T^{-\frac{1}{2}}\sum_{t=1}^{T} \bigtriangledown_{H'F}F \Big(-((H'F_{t-1})'-(\overline{H'F}_{t-1})')\hat{\beta}(\tau)\Big)*(\hat{F}_{t-1}-H'F_{t-1}) exp(\xi'\Phi(I_{t-1}))}_{\text{Term II\textsubscript{1}}}\\
&\quad \quad + \underbrace{T^{-\frac{1}{2}}\sum_{t=1}^{T} \bigtriangledown_{H'F}F \Big(-((H'F_{t-1})'-(\overline{H'F}_{t-1})')\hat{\beta}(\tau)\Big)(\hat{F}_{t-1}-H'F_{t-1}) \bigtriangledown_{H'F}exp(\xi'\Phi(\overline{I}_{t-1}))*(\hat{F}_{t-1}-H'F_{t-1})}_{\text{Term II\textsubscript{2}}}\\
\end{split}
\end{align*}
\\
Focusing therefore on \textbf{Term II\textsubscript{1}} employing the factor estimation error identity:
\begin{align*}
\begin{split}
II_2&=T^{-\frac{1}{2}}\sum_{t=1}^{T} \bigtriangledown_{H'F} F \Big(-((H'F_{t-1})'-(\overline{H'F}_{t-1})')\hat{\beta}(\tau)\Big)*(\hat{F}_{t-1}-H'F_{t-1}) \Big]exp(b'\Phi(I_{t-1}))\\
& \leq  \sup_{t} \Big( exp(\xi'\Phi(I_{t-1}))\Big) *\Big[T^{-\frac{1}{2}}\sum_{t=1}^{T} \underbrace{\bigtriangledown_{H'F}F \Big(-((H'F_{t-1})'-(\overline{H'F}_{t-1})')\hat{\beta}(\tau)\Big)}_{W_{t-1}}*(\hat{F}_{t-1}-H'F_{t-1}) \Big]\\
&=\sup_{t} \Big( exp(\xi'\Phi(I_{t-1}))\Big)* T^{-\frac{1}{2}}\sum_{t=1}^{T}  W_{t-1}*\Big\lbrace V_{NT}^{-1}\big[\frac{1}{T}\sum_{s=1}^{T}\hat{F}_s\gamma_N(s,t)+\frac{1}{T}\sum_{s=1}^{T}\hat{F}_s\zeta_{st}+\frac{1}{T}\sum_{s=1}^{T}\hat{F}_s\eta_{st}+\frac{1}{T}\sum_{s=1}^{T}\hat{F}_s\xi_{st}\big] \Big\rbrace\\
&= \sup_{t} \Big( exp(\xi'\Phi(I_{t-1}))\Big)*V^{-1}_{NT} \Big\lbrace T^{-\frac{3}{2}}\sum_{t=1}^{T}  W_{t-1}\hat{F}_s\gamma_N(s,t)+T^{-\frac{3}{2}}\sum_{t=1}^{T}  W_{t-1}\hat{F}_s\zeta_{st}+T^{-\frac{3}{2}}\sum_{t=1}^{T}  W_{t-1}\hat{F}_s\eta_{st}\\
&\quad \quad \quad \quad \quad \quad \quad \quad \quad \quad \quad \quad \quad \quad \quad \quad \quad \quad \quad \quad \quad \quad \quad \quad \quad \quad \quad \quad \quad \quad \quad \quad \quad +T^{-\frac{3}{2}}\sum_{t=1}^{T} W_{t-1}\hat{F}_s\xi_{st} \Big\rbrace \\ 
\end{split}
\end{align*}
where $E(\Vert W_{t-1} \Vert) \leq M$ and $E(\Vert W_{t-1} \Vert^2) \leq M$ by construction.\\ 

Taking the first term of \textbf{Term II\textsubscript{1}}, once again neglecting $V^{-1}_{NT}$ and $\sup_{t} \Big( exp(\xi'\Phi(I_{t-1}))\Big) $ which are bounded by definition, then by adding and subtracting terms we obtain the following:
\begin{align*}
\begin{split}
T^{-\frac{3}{2}}\sum_{t=1}^{T}\sum_{s=1}^{T}  W_{t-1}\hat{F}_s\gamma_N(s,t)=T^{-\frac{3}{2}}\sum_{t=1}^{T} \sum_{s=1}^{T} W_{t-1}H'F_s\gamma_N(s,t)+T^{-\frac{3}{2}}\sum_{t=1}^{T} \sum_{s=1}^{T} W_{t-1}(\hat{F}_s-H'F_s)\gamma_N(s,t)
\end{split}
\end{align*}\\

The first component, under assumptions A and C2 is bounded by:
\begin{align*}
\begin{split}
E[T^{-\frac{3}{2}}\sum_{t=1}^{T}\sum_{s=1}^{T}  W_{t-1}H'F_s\gamma_N(s,t)]&=T^{-\frac{3}{2}}\sum_{t=1}^{T} \sum_{s=1}^{T} E[W_{t-1}H'F_s\gamma_N(s,t)]\\
& \leq T^{-\frac{3}{2}}\sum_{t=1}^{T} \sum_{s=1}^{T} (E \Vert W_{t-1} \Vert^2)^{\frac{1}{2}}(E \Vert F_s \Vert^2)^{\frac{1}{2}} \vert \gamma_N(s,t) \vert\\
&=M^2*T^{-\frac{1}{2}}* T^{-1}\sum_{t=1}^{T}\sum_{s=1}^{T} \vert \gamma_N(s,t) \vert\\
&=O_p(\frac{1}{\sqrt{T}})
\end{split}
\end{align*}

Meanwhile, under lemma A.2 and lemma 1i in \citet{bai_determining_2002}:
\begin{align*}
\begin{split}
T^{-\frac{3}{2}}\sum_{t=1}^{T} \sum_{s=1}^{T} W_{t-1}(\hat{F}_s-H'F_s)\gamma_N(s,t)&\leq (T^{-1}\sum_{s=1}^{T} \Vert \hat{F}_s-H'F_s \Vert^2(T^{-1}\sum_{t=1}^T \Vert W_{t-1} \Vert^2)^{\frac{1}{2}} (T^{-1}\sum_{t=1}^{T}\sum_{s=1}^{T} \vert \gamma_N(s,t) \vert^2)^{\frac{1}{2}}\\
&=O_p(\frac{1}{\delta_{NT}\sqrt{T}})
\end{split}
\end{align*}
Therefore the first term is an $O_p(\frac{1}{\sqrt{T}})$.\\
\\
\\
Focusing now on the second term, with a similar manipulation:\\
\begin{align*}
\begin{split}
T^{-\frac{3}{2}}\sum_{t=1}^{T} \sum_{s=1}^{T} W_{t-1}\hat{F}_s\zeta_{st}=T^{-\frac{3}{2}}\sum_{t=1}^{T}\sum_{s=1}^{T}  W_{t-1}H'F_s\zeta_{st}+T^{-\frac{3}{2}}\sum_{t=1}^{T} \sum_{s=1}^{T} W_{t-1}(\hat{F}_s-H'F_s)\zeta_{st}
\end{split}
\end{align*}\\

Under assumption F1, the first component is bounded by:
\begin{align*}
\begin{split}
T^{-\frac{3}{2}}\sum_{t=1}^{T}\sum_{s=1}^{T}  W_{t-1}H'F_s\zeta_{st}&=\sum_{t=1}^{T} \sum_{s=1}^{T} W_{t-1}F_s \Big( \frac{1}{N} \sum_{i=1}^{N} e_{is}e_{it}-\gamma_N(s,t) \Big)\\
& = H' *\frac{1}{\sqrt{N}} T^{-1}\sum_{t=1}^{T}  W_{t-1}\frac{1}{\sqrt{NT}}\sum_{s=1}^{T} \sum_{i=1}^{N} F_s( e_{is}e_{it}-\gamma_N(s,t))\\
& \leq H'*\frac{1}{\sqrt{N}}* \Big( T^{-1}\sum_{t=1}^{T} \Vert W_{t-1}\Vert^2 \Big)^{\frac{1}{2}} \Big[ T^{-1}\sum_{t=1}^{T} \Big( \frac{1}{\sqrt{NT}}\sum_{s=1}^{T} \sum_{i=1}^{N} F_s(e_{is}e_{it}-\gamma_N(s,t)) \Big)^2 \Big]^{\frac{1}{2}}\\
&= O_p(\frac{1}{\sqrt{N}})
\end{split}
\end{align*}\\

The second component, under lemma A.2 and assumption F1 is bounded by:
\begin{align*}
\begin{split}
T^{-\frac{3}{2}}\sum_{t=1}^{T} \sum_{s=1}^{T} W_{t-1}(\hat{F}_s-H'F_s)\zeta_{st} &=T^{-\frac{3}{2}}\sum_{t=1}^{T} \sum_{s=1}^{T} W_{t-1}(\hat{F}_s-H'F_s)\Big( \frac{1}{N} \sum_{i=1}^{N} e_{is}e_{it}-\gamma_N(s,t) \Big)\\
&\leq \frac{\sqrt{T}}{\sqrt{N}}*(T^{-1}\sum_{s=1}^{T} \Vert \hat{F}_s-H'F_s \Vert^2)^{\frac{1}{2}}\Big[T^{-1}\sum_{S=1}^{T} \Big( \frac{1}{\sqrt{NT}}\sum_{T=1}^{T} \sum_{i=1}^{N} W_{t-1}(e_{is}e_{it}-\gamma_N(s,t)) \Big)^2 \Big]^{\frac{1}{2}}\\
&\leq \frac{\sqrt{T}}{\sqrt{N}}*(T^{-1}\sum_{s=1}^{T} \Vert \hat{F}_s-H'F_s \Vert^2)^{\frac{1}{2}}\Big[ T^{-1}\sum_{s=1}^{T} \Big( \frac{1}{T}\sum_{t=1}^{T} (\frac{1}{\sqrt{N}}\sum_{i=1}^{N} (e_{is}e_{it}-\gamma_N(s,t)))W_{t-1} \Big)^2 \Big]^{\frac{1}{2}}\\
&=O_p(\frac{\sqrt{T}}{\delta_{NT}\sqrt{N}})
\end{split}
\end{align*}\\
Therefore, the second term is an $O_p(\frac{1}{\sqrt{N}})$.\\
\\
\\
Similarly,for the third term of \textbf{Term II\textsubscript{1}}, under lemma A.2 and assumption F3, we obtain that:
\begin{align*}
\begin{split}
T^{-\frac{3}{2}}\sum_{t=1}^{T}\sum_{s=1}^{T}  W_{t-1}H'F_s\eta_{st}&=\sum_{t=1}^{T} \sum_{s=1}^{T} W_{t-1}H'F_s \Big( \frac{1}{N} \sum_{i=1}^{N} \lambda_i F_s'e_{it}\Big)\\
& =  \Big (T^{-1}\sum_{s=1}^T F_sF_s' \Big) \frac{1}{\sqrt{T}N}\sum_{t=1}^T\sum_{i=1}^N
W_{t-1}\lambda_i e_{it}\\
& \leq H' \Big (T^{-1}\sum_{s=1}^T F_sF_s' \Big) \frac{\sqrt{T}}{\sqrt{N}} \Big(T^{-1}\sum_{t=1}^T W_{t-1} ( \frac{1}{\sqrt{N}}\sum_{i=1}^N\lambda_i e_{it}\Big)\\
&= O_p(\frac{\sqrt{T}}{\sqrt{N}})
\end{split}
\end{align*}

\begin{align*}
\begin{split}
T^{-\frac{3}{2}}\sum_{t=1}^{T} \sum_{s=1}^{T} W_{t-1}(\hat{F}_s-H'F_s)\eta_{st} &=  T^{-\frac{3}{2}}\sum_{t=1}^{T} \sum_{s=1}^{T} W_{t-1}(\hat{F}_s-H'F_s)\Big(\frac{F_s\Lambda e_t}{N}\Big)\\
&= \Big(T^{-1}\sum_{s=1}^{T} (\hat{F}_s-H'F_s)F_s \Big)\Big(\frac{1}{NT} \sum_{t=1}^{T}W_{t-1}\Lambda e_{t} \Big)*\sqrt{T}\\
& \leq \Big(T^{-1}\sum_{s=1}^{T}  \Vert \hat{F}_s-H'F_s \Vert^2\Big)^{\frac{1}{2}}\Big(T^{-1}\sum_{s=1}^{T}  \Vert F_s \Vert^2 \Big)^{\frac{1}{2}}\Big(T^{-1}\sum_{t=1}^T W_{t-1} ( \frac{1}{\sqrt{N}}\sum_{i=1}^N\lambda_i e_{it}\Big)*\frac{\sqrt{T}}{\sqrt{N}}\\
&=O_p(\frac{\sqrt{T}}{\delta_{NT}\sqrt{N}})
\end{split}
\end{align*}\\
Thus the third term is an $O_p(\frac{\sqrt{T}}{\sqrt{N}})$.\\

Lastly, for the fourth term of \textbf{Term II\textsubscript{1}}, by lemma A.2 and assumptions A and F3, we obtain:
\begin{align*}
\begin{split}
T^{-\frac{3}{2}}\sum_{t=1}^{T}\sum_{s=1}^{T}  W_{t-1}H'F_s\xi_{st}&=\sum_{t=1}^{T} \sum_{s=1}^{T} W_{t-1}H'F_s \Big( \frac{F_t' \Lambda' e_s}{N}\Big)\\
& = H' \Big (T^{-1}\sum_{s=1}^T \frac{F_s e_s' \Lambda}{N} \Big) \Big( T^{-1}\sum_{t=1}^T W_{t-1}F_t' \Big)* \sqrt{T}\\
& \leq \frac{\sqrt{T}}{\sqrt{N}}* \Big(T^{-1}\sum_{s=1}^{T}  \Vert F_s \Vert^2 \Big)^{\frac{1}{2}} \Big(T^{-1} \sum_{t=1}^{T} \Vert \frac{\Lambda e_{t}}{\sqrt{N}} \Vert^2\Big)^{\frac{1}{2}}\Big(\frac{1}{T} \sum_{t=1}^{T} \Vert W_{t-1} \Vert^2 \Big)^{\frac{1}{2}}\Big(T^{-1}\sum_{t=1}^{T}  \Vert F_t \Vert^2 \Big)^{\frac{1}{2}}\\
&= O_p(\frac{\sqrt{T}}{\sqrt{N}})
\end{split}
\end{align*}

\begin{align*}
\begin{split}
T^{-\frac{3}{2}}\sum_{t=1}^{T} \sum_{s=1}^{T} W_{t-1}(\hat{F}_s-H'F_s)\xi_{st} &=  T^{-\frac{3}{2}}\sum_{t=1}^{T} \sum_{s=1}^{T} W_{t-1}(\hat{F}_s-H'F_s)\Big( \frac{F_t' \Lambda' e_s}{N}\Big)\\
&= \Big[T^{-1}\sum_{s=1}^{T} (\hat{F}_s-H'F_s)\frac{\Lambda' e_s}{N} \Big]\Big[T^{-1} \sum_{t=1}^{T}W_{t-1}F_t \Big]*\sqrt{T}\\
& \leq \frac{\sqrt{T}}{\sqrt{N}}*\Big(T^{-1}\sum_{s=1}^{T}  \Vert \hat{F}_s-H'F_s \Vert^2\Big)^{\frac{1}{2}}*\Big(T^{-1}\sum_{s=1}^{T}  \Vert \frac{\Lambda' e_s}{N} \Vert^2 \Big)^{\frac{1}{2}}\Big(\frac{1}{T} \sum_{t=1}^{T} \Vert W_{t-1} \Vert^2 \Big)^{\frac{1}{2}}\\
& \quad \quad \quad \quad \quad \quad \quad \quad \quad \quad \quad \quad \quad \quad \quad \quad \quad \quad \quad \quad \quad \quad \quad \quad \Big(\underbrace{T^{-1}\sum_{t=1}^{T}  \Vert F_t \Vert^2 }_{O_p(1)}\Big)^{\frac{1}{2}}\\
&=O_p(\frac{\sqrt{T}}{\delta_{NT}\sqrt{N}})
\end{split}
\end{align*}\\

Thus the last term is an $O_p(\frac{\sqrt{T}}{\sqrt{N}})$.\\

Overall therefore,
\begin{align*}
\textbf{Term II\textsubscript{1}}&=T^{-\frac{1}{2}}\sum_{t=1}^{T} \bigtriangledown_{H'F} F \Big(-((H'F)'_{t-1}-(\overline{H'F})'_{t-1})\hat{\beta}(\tau)\Big)*(\hat{F}_{t-1}-H'F_{t-1}) \Big]exp(\xi'\Phi(I_{t-1}))\\
&=O_p(\frac{1}{\sqrt{T}})+O_p(\frac{\sqrt{T}}{\delta_{NT}\sqrt{N}})+O_p(\frac{\sqrt{T}}{\sqrt{N}})+O_p(\frac{\sqrt{T}}{\sqrt{N}})\\
&=O_p\Big(\frac{\sqrt{T}}{\min\lbrace\sqrt{N},T\rbrace}\Big)
\end{align*}\\
\\

Now focusing on \textbf{Term II\textsubscript{2}}. Because $(x+y+z+u)^2 \leq 4(x^2+y^2+z^2+u^2)$:
\begin{align*}
\begin{split}
II_2&=T^{-\frac{1}{2}}\sum_{t=1}^{T} \bigtriangledown_{H'F}F \Big(-((H'F)'_{t-1}-(\overline{H'F})'_{t-1})\hat{\beta}(\tau)\Big)(\hat{F}_{t-1}-H'F_{t-1}) \bigtriangledown_{H'F}exp(\xi'\Phi(\overline{I}_{t-1}))*(\hat{F}_{t-1}-H'F_{t-1})\\
&\leq \sup_{t} \Big[\bigtriangledown_{H'F}exp(\xi'\Phi(\overline{I}_{t-1}))\Big]T^{-\frac{1}{2}}\sum_{t=1}^{T} \bigtriangledown_{H'F}F \Big(-((H'F)'_{t-1}-(\overline{H'F})'_{t-1})\hat{\beta}(\tau)\Big)(\hat{F}_{t-1}-H'F_{t-1})' *(\hat{F}_{t-1}-H'F_{t-1})\\
&\leq \sup_{t} \Big[\bigtriangledown_{H'F}exp(\xi'\Phi(\overline{I}_{t-1}))\Big]T^{-\frac{1}{2}}\sum_{t=1}^{T} \Vert W_{t-1}\Vert \Vert \hat{F}_{t-1}-H'F_{t-1}\Vert^2\\
&\leq M*T^{-\frac{1}{2}}\sum_{t=1}^{T} \Big \lbrace \Vert W_{t-1}\Vert* 4 \Big [ T^{-2}\Vert \sum_{s=1}^T \hat{F}_s\gamma_N{s,t}\Vert^2+T^{-2}\Vert \sum_{s=1}^T \hat{F}_s\zeta_{st}\Vert^2+T^{-2}\Vert \sum_{s=1}^T \hat{F}_s\eta_{st}\Vert^2+T^{-2}\Vert \sum_{s=1}^T \hat{F}_s\xi_{st}\Vert^2 \Big] \text{\footnote{$(a+b+c+d)^2 \leq 4(a^2+b^2+c^2+d^2)$}}\Big \rbrace\\
&\leq 4M*T^{-\frac{5}{2}}\sum_{t=1}^{T} \Big \lbrace \Vert W_{t-1}\Vert*  \Big [\Vert \sum_{s=1}^T \hat{F}_s\gamma_N{s,t}\Vert^2+\Vert \sum_{s=1}^T \hat{F}_s\zeta_{st}\Vert^2+\Vert \sum_{s=1}^T \hat{F}_s\eta_{st}\Vert^2+\Vert \sum_{s=1}^T \hat{F}_s\xi_{st}\Vert^2 \Big]\Big \rbrace\\
\end{split}
\end{align*}\\
\\

Focusing on the first term of \textbf{Term II\textsubscript{2}}, following from the fact that $T^{-1} \sum_{s=1}^T \Vert \hat{F}_s\Vert^2=O_p(1)$ and under assumptions E1, we obtain:
\begin{align*}
\begin{split}
T^{-\frac{5}{2}}\sum_{t=1}^{T}  \Vert W_{t-1}\Vert*  \Big[ \Vert \sum_{s=1}^T \hat{F}_s\gamma_N{s,t}\Vert^2 \Big] &\leq T^{-\frac{3}{2}} \Big( T^{-1}\sum_{t=1}^{T}  \Vert W_{t-1}\Vert^2 \Big)^{\frac{1}{2}}  \Big( T^{-1}\sum_{t=1}^T \Big[\Vert \sum_{s=1}^T \hat{F}_s\gamma_N{s,t}\Vert^2 \Big]^2 \Big)^{\frac{1}{2}} \\
& \leq T^{-\frac{1}{2}} \Big( T^{-1}\sum_{t=1}^{T}  \Vert W_{t-1}\Vert^2\Big)^{\frac{1}{2}} \Big( T^{-1}\sum_{t=1}^T \Big[  \Big(T^{-1} \sum_{s=1}^T \Vert \hat{F}_s\Vert^2 \Big)^{\frac{1}{2}} \Big(\sum_{s=1}^T \vert \gamma_N{s,t} \vert^2 \Big)^{\frac{1}{2}} \Big]^4 \Big)^{\frac{1}{2}} \\
&=O_p \Big(\frac{1}{\sqrt{T}}\Big)
\end{split}
\end{align*}\\

Meanwhile, the second term of \textbf{Term II\textsubscript{2}}, under assumption C5 is bounded by:
\begin{align*}
\begin{split}
T^{-\frac{5}{2}}\sum_{t=1}^{T}  \Vert W_{t-1}\Vert* & \Big[ \Vert \sum_{s=1}^T \hat{F}_s \zeta_{st} \Vert^2 \Big]\\
&\leq T^{-\frac{3}{2}} \Big( T^{-1}\sum_{t=1}^{T}  \Vert W_{t-1}\Vert^2 \Big)^{\frac{1}{2}}  \Big( T^{-1}\sum_{t=1}^T \Big[\Vert \sum_{s=1}^T \hat{F}_s\zeta_{st}\Vert^2 \Big]^2 \Big)^{\frac{1}{2}} \\
& \leq \sqrt{T} \Big(T^{-1}\sum_{t=1}^{T}  \Vert W_{t-1}\Vert^2 \Big)^{\frac{1}{2}}  \Big( T^{-1}\sum_{t=1}^T \Big[  \Big(T^{-1} \sum_{s=1}^T \Vert \hat{F}_s\Vert^2 \Big)^{\frac{1}{2}} \Big(T^{-1}\sum_{s=1}^T \Vert \zeta_{st} \Vert^2 \Big)^{\frac{1}{2}} \Big]^4 \Big)^{\frac{1}{2}} \\
& \leq \sqrt{T} \Big( T^{-1}\sum_{t=1}^{T}  \Vert W_{t-1}\Vert^2 \Big)^{\frac{1}{2}}  \Big( T^{-1}\sum_{t=1}^T \Big[  \Big(T^{-1} \sum_{s=1}^T \Vert \hat{F}_s\Vert^2\Big)^{\frac{1}{2}} \Big(T^{-1}\sum_{s=1}^T \vert N^{-1}\sum_{i=1}^N e_{is}e_{it}-E(e_{is}e_{it})  \vert^2 \Big)^{\frac{1}{2}} \Big]^4 \Big)^{\frac{1}{2}} \\
&=O_p \Big(\frac{\sqrt{T}}{\sqrt{N}}\Big)
\end{split}
\end{align*}\\

The third term under assumption F3 is bounded by:
\begin{align*}
\begin{split}
T^{-\frac{5}{2}}\sum_{t=1}^{T}  \Vert W_{t-1}\Vert*  \Big[ \Vert \sum_{s=1}^T \hat{F}_s\gamma_N{s,t}\Vert^2 \Big] &\leq T^{-\frac{3}{2}} \Big( T^{-1}\sum_{t=1}^{T}  \Vert W_{t-1}\Vert^2 \Big)^{\frac{1}{2}}  \Big( T^{-1}\sum_{t=1}^T \Big[\Vert \sum_{s=1}^T \hat{F}_s\eta_{st}\Vert^2 \Big]^2 \Big)^{\frac{1}{2}} \\
& \leq T^{-\frac{1}{2}} \Big(T^{-1}\sum_{t=1}^{T}  \Vert W_{t-1}\Vert^2 \Big)^{\frac{1}{2}}  \Big( T^{-1}\sum_{t=1}^T \Big[  \Big(T^{-1} \sum_{s=1}^T \Vert \hat{F}_s\Vert^2 \Big)^{\frac{1}{2}} \Big(\sum_{s=1}^T \Vert \eta_{st} \Vert^2 \Big)^{\frac{1}{2}} \Big]^4 \Big)^{\frac{1}{2}} \\
&=O_p \Big(\frac{\sqrt{T}}{N}\Big)
\end{split}
\end{align*}\\

Similarly, the fourth term of \textbf{Term II\textsubscript{2}} is bounded by:
\begin{align*}
\begin{split}
T^{-\frac{5}{2}}\sum_{t=1}^{T}  \Vert W_{t-1}\Vert* & \Big[ \Vert \sum_{s=1}^T \hat{F}_s \xi_{st} \Vert^2 \Big]\\
&\leq T^{-\frac{3}{2}} \Big( T^{-1}\sum_{t=1}^{T}  \Vert W_{t-1}\Vert^2 \Big)^{\frac{1}{2}}  \Big( T^{-1}\sum_{t=1}^T \Big[\Vert \sum_{s=1}^T \hat{F}_s\xi_{st}\Vert^2 \Big]^2 \Big)^{\frac{1}{2}} \\
& \leq \sqrt{T} \Big(T^{-1}\sum_{t=1}^{T}  \Vert W_{t-1}\Vert^2 \Big)^{\frac{1}{2}}  \Big( T^{-1}\sum_{t=1}^T \Big[  \Big(T^{-1} \sum_{s=1}^T \Vert \hat{F}_s\Vert^2 \Big)^{\frac{1}{2}} \Big(T^{-1}\sum_{s=1}^T \Vert \xi_{st}\Vert^2 \Big)^{\frac{1}{2}} \Big]^4 \Big)^{\frac{1}{2}} \\
& \leq \sqrt{T} \Big(T^{-1}\sum_{t=1}^{T}  \Vert W_{t-1}\Vert^2 \Big)^{\frac{1}{2}}  \Big( T^{-1}\sum_{t=1}^T \Big[  \Big(T^{-1} \sum_{s=1}^T \Vert \hat{F}_s\Vert^2 \Big)^{\frac{1}{2}} \Big(T^{-1}\sum_{s=1}^T \Vert \frac{F_t'\Lambda'e_s}{N}\Vert^2 \Big)^{\frac{1}{2}} \Big]^4 \Big)^{\frac{1}{2}} \\
& \leq \sqrt{T} \Big( T^{-1}\sum_{t=1}^{T}  \Vert W_{t-1}\Vert^2 \Big)^{\frac{1}{2}}  \Big( T^{-1}\sum_{t=1}^T \Big[  \Big(T^{-1} \sum_{s=1}^T \Vert \hat{F}_s\Vert^2 \Big)^{\frac{1}{2}} \Big(\Vert F_t \Vert^2* N^{-1}T^{-1}\sum_{s=1}^T \Vert \frac{\Lambda' e_s}{\sqrt{N}} \Vert^2 \Big)^{\frac{1}{2}} \Big]^4 \Big)^{\frac{1}{2}} \\
&=O_p \Big(\frac{\sqrt{T}}{N}\Big)
\end{split}
\end{align*}\\

Overall therefore,
\begin{align*}
\textbf{Term II\textsubscript{2}}&=T^{-\frac{1}{2}}\sum_{t=1}^{T} \bigtriangledown_{H'F}F \Big(-((H'F)'_{t-1}-(\overline{H'F})'_{t-1})\hat{\beta}(\tau)\Big)(\hat{F}_{t-1}-H'F_{t-1}) \bigtriangledown_{H'F}exp(\xi'\Phi(\overline{I}_{t-1}))*(\hat{F}_{t-1}-H'F_{t-1})\\
&=O_p(\frac{1}{\sqrt{T}})+O_p(\frac{\sqrt{T}}{\sqrt{N}})+O_p(\frac{\sqrt{T}}{N})+O_p(\frac{\sqrt{T}}{N})\\
&=O_p\Big(\frac{\sqrt{T}}{\min\lbrace\sqrt{N},T\rbrace}\Big).
\end{align*}\\

This concludes the proof which demonstrates that factor estimation error does not influence the test statistic, since uniformly in $\tau$:\\
\begin{align*}
S_{2,T}(\xi,\hat{\theta}(\tau))&=T^{-\frac{1}{2}}\sum_{t=1}^{T} [\mathbbm{1}(\tilde{\epsilon}_t\leq 0)-\tau]exp(\xi'\Phi(\hat{I}_{t-1}))\\
&=T^{-\frac{1}{2}}\sum_{t=1}^{T} [\mathbbm{1}(\hat{\epsilon}_t\leq 0)-\tau]exp(\xi'\Phi(I_{t-1}))+ \boldsymbol{I_2}+\boldsymbol{II_1}+\boldsymbol{II_2}\\
&=T^{-\frac{1}{2}}\sum_{t=1}^{T} [\mathbbm{1}(\hat{\epsilon}_t\leq 0)-\tau]exp(\xi'\Phi(I_{t-1}))+O_p\Big(\frac{\sqrt{T}}{\min\lbrace\sqrt{N},T\rbrace}\Big)\\
&=T^{-\frac{1}{2}}\sum_{t=1}^{T} [\mathbbm{1}(\hat{\epsilon}_t\leq 0)-\tau]exp(\xi'\Phi(I_{t-1}))+o_p(1) 
\end{align*}

\vspace{1cm}

\textbf{Proof of Theorem 1.} 
This theorem establishes the limit distribution of $S_{j,T}(\xi, \hat{\theta}(\tau))$.  Under the null hypothesis $H_{0,j}$ and Assumptions  A-F, factor estimation error is negligible by Lemma 1. The theorem then follows from the fact under the additional assumptions  G-J,
\begin{align*}
\sup\limits_{\xi \in \Upsilon,\tau \in \mathcal{T}}\vert &S_{j,T}(\xi, \hat{\theta}_T(\tau))-S_{j,T}(\xi,  \theta(\tau))+G'(\xi, \theta(\tau))Q(\tau) \vert=o_p(1).
\end{align*}
The statement demonstrates that the quantile-marked empirical process converges to a test statistic where parameter estimation error is not present and a component accounting for the fact that in practice $\theta(\tau)$ is unknown and has to be estimated from a sample $\lbrace (Y_t, I'_{t-1})':1 \leq t \leq T \rbrace$ by an estimator $\hat{\theta}_T(\tau)$.  See Theorem 2 in \citet{escanciano_specification_2010}.\\

Under the null hypothesis and assumption G1, given that $S_{j,T}(\xi, \theta(\tau))$ is a zero-mean-square-integrable martingale for each $\nu=(\xi', \tau)' \in \Pi$, with $\Pi=\Upsilon \times \mathcal{T}$, the finite dimensional distributions of $S_{j,T}(\xi, \theta(\tau))$ weakly converge in the space $l^{\infty}(\Pi)$ to $S_{j,\infty}(\xi, \theta(\tau))$, a multivariate normal distribution with a zero mean vector and a covariance function given by:
\begin{align}
Cov_{\infty}(\nu_1, \nu_2)=(\min \lbrace \tau_1,\tau_2 \rbrace-\tau_1\tau_2) E[exp((\xi_1-\xi_2)'\Phi(I_0))] \label{Statistic Covariance function}
\end{align}
where, $\nu_1=(\xi_1',\tau_1)'$ and $\nu_2=(\xi_2',\tau_2)'$ are generic elements of $\Upsilon \times \mathcal{T}$ (See Theorem 1 in \citet{escanciano_specification_2010})). It is immediate to see here that the covariance function does not contain any time dependent cross terms, as under the null $\mathbbm{1}(y_t-m_j(I_{t-1},\hat{\theta}(\tau))\leq 0)-\tau$ is a martingale difference sequence.\\ 

Given also that under Assumption J, the quantile limit process $Q(\tau)$ is a zero mean Gaussian Process with a covariance function given by equation~(\ref{Quantile Process Covariance Function}), the result then follows from the interaction of two separate Gaussian Processes.

\newpage
\subsection{Dataset} 
\begin{longtable}{|l|l|l|}
\hline
\textbf{NR} & \textbf{FAME CODE} & \textbf{Series Name} \\
\hline
1&D7BT&Consumer Price Index: all items\\
2&ABMI&Gross Domestic Product: chained volume measures\\
&&\\
3&ABJR&Household final consumption expenditure \\
4&CKYY&IOP: Industry D: Manufacturing\\
5&CKYZ&IOP: Industry E: Electricity, gas and water supply\\
6&CKZF&IOP: Industry DF: Manufacturing of food, drink and tobacco\\
7&CKZG&IOP: Industry DG: Manufacturing of chemicals and man-made fibres \\
8&GDBQ&ESA95 Output Index: F:Construction\\
9&GDQH&SA95 Output Industry: I: Transport storage and communication\\
10&GDQS&SA95 Output Industry: G-Q: Total\\
11&IKBK&Balance of Payments: Trade in Goods and Services: Total exports\\
12&IKBL&Balance of Payments: Imports: Total Trade in Goods and Services\\
13&NMRY&General Government: Final consumption expenditure\\
14&NPQT&Total Gross Fixed Capital Formation\\ \hline
\multicolumn{3}{c}{ \textbf{Household final consumption expenditure: durable goods}}\\ \hline
15&ATQX&Furniture and households\\
16&ATRD&Carpets and other floor coverings\\
17&ATRR&Telephone and telefax equipment\\
18&ATRV&Audio visual equipment\\
19&ATRZ&Photo and cinema equipment and optical instruments\\
20&ATSD&Information processing equipment\\
21&LLKX&All funrishing and household\\
22&LLKY&All health\\
23&LLKZ&All transport\\
24&LLLA&All communication\\
25&LLLB&All recreation and culture\\
26&LLLC&All miscellaneous\\
27&TMMI&All purchases of vehicles\\
28&TMML&Motor cars\\
29&TMMZ&Motor cycles\\
30&TMNB&Major durables for outdoor recreation\\
31&TMNO&Bicycles\\
32&UTID&Total \\
33&UWIC&Therapeutic appliances and equipment\\
34&XYJP&Major house appliances\\
35&XYJR&Major tools and equipment\\
36&XYJT&Musical instruments and major durables for indoor recreation\\
37&ZAYM&Jewelery, clocks and watches\\ \hline
\multicolumn{3}{c}{ \textbf{Household final consumption expenditure: semi-durable goods}}\\ \hline
38&ATQV&Shoes and other footwear\\
39&ATRF&Household and textiles\\
40&ATRJ&Glassware, tableware and household utensils\\
41&ATSH&Recording media\\
42&ATSL&Games, toys and hobbies\\
43&ATSX&Other personal effects\\
44&AWUW&Motor vehicle spares\\
45&CDZQ&Books\\
46&LLLZ&All clothing and footwear\\
47&LLMC&All recreation and culture\\
48&LLMD&All miscellaneous\\
49&UTIT&Total  \\
50&XYJN&Clothing materials\\
51&XYJO&Other articles of clothing and clothing accessories\\
52&XYJQ&Small eectric household appliances\\
53&XYJS&Small tools and miscellaneous accessories\\
54&XYJU&Equipment for sport, camping etc\\
55&XYJX&Electrical appliances for personal care\\
56&ZAVK&Garments\\ \hline
\multicolumn{3}{c}{ \textbf{Household final consumption expenditure: non-durable goods}}\\ \hline
57&ATSP&Other products for personal care\\
58&ATUA&Materials for the maintenance and repair of the dwelling\\
59&AWUX&Gardens, plants and flowers\\
60&CCTK&Meat\\
61&CCTL&Fish\\
62&CCTM&Milk, cheese and eggs\\
63&CCTN&Oils and fats\\
64&CCTO&Fruit\\
65&CCTT&Coffee,tea and cocoa\\
66&CCTU&Mineral, water and soft drinks\\
67&CCTY&Vehicle fuels and lubricants\\
68&CCUA&Electricity\\
69&CDZY&Newspapers and periodicals\\
70&LLLL&All housing, water, electricity, gas and other fuels\\
71&LLLM&All furnishing and household goods\\
72&LLLN&All health\\
73&LLLO&All transport\\
74&LLLP&All recreation and culture\\
75&LLLQ&All miscellaneous\\
76&LTZA&Gas\\
77&LTZC&Liquid fuels\\
78&TTAB&Solid fuels\\
79&UTHW&Wine, cider and sherry\\
80&UTIL&Total\\
81&UTXP&Pharmaceutical Products\\
82&UTZN&Water supply\\
83&UUIS&Spirits\\
84&UUVG&Beer\\
85&UWBK&All food \\
86&UWBL&Bread and cereals\\
87&UWFD&Vegetables\\
88&UWFX&Sugar and sweet products\\
89&UWGH&Food products n.e.c.\\
90&UWGI&All non-alcoholic beverages\\
91&UWHO&Non-durable household goods\\
92&UWIB&Other medical products\\
93&UWKQ&Pets and related products\\
94&XYJV&Miscellaneous printed matter\\
95&XYJW&Stationary and drawing materials\\
96&ZAKY&All alcoholic beverages and otbacco\\
97&ZWUN&All food and non-alcoholic beverages\\
98&ZWUP&Tobacco\\
99&ZWUR&All elctricity, gas and other fuels\\ \hline
\multicolumn{3}{c}{ \textbf{Household final consumption expenditure: services}}\\ \hline
100&AWUY&Repair and hire of footwear\\
101&AWUZ&Services for the maintenance and reair of the dwelling\\
102&AWVA&Vehicle maintenance and repair\\
103&AWVB&Railways\\
104&AWVC&Air\\
105&AWVD&Sea and inland waterway\\
106&AWVE&Other\\
107&CCUO&Imputed rentals of owner-occupiers\\
108&CCVA&Games of chance\\
109&CCVM&Postal services\\
110&CCVZ&Hairdressing salons and personal grooming establishments\\
111&GBFG&Actual rentals paid by tenants\\
112&GBFK&All imputed rentals for housing\\
113&GBFN&Other imputed rentals\\
114&LLLR&All clothing and footwear\\
115&LLLS&All housing, water, electricity, gas and other fuels\\
116&LLLT&All funrishing and household\\
117&LLLU&All health\\
118&LLLV&Total transport\\
119&LLLW&All communication\\
120&LLLX&All recreation and culture\\
121&LLLY&All miscellaneous\\
122&UTIP&Total\\
123&UTMH&Paramedical services\\
124&UTYF&Hospital services\\
125&UTYH&Life insurance \\
126&UTZX&Sewerage collection\\
127&UWHI&Clothing, repair and hire of clothing\\
128&UWHK&Refuse collection\\
129&UWHM&Repair of furniture, furnishings and floor coverings\\
130&UWHN&Repair of household appliances\\
131&UWIA&Domestic and household services\\
132&UWKO&Repair of audio-visual, ohoto and information processing equipment\\
133&UWKP&Maintenance of other major durables for recreation and culture\\
134&UWLD&Veterinary and other services\\
135&ZAVQ&All actual rentals for housing \\
136&ZAWG&All out-patient services\\
137&ZAWI&Medical services\\
138&ZAWK&Dental services\\
139&ZAWQ&Other vehicle services\\
140&ZAWS&All transport services\\
141&ZAWU&Road\\
142&ZAWY&Telephone and telefax services\\
143&ZAXI&All recreational and cultural services\\
144&ZAXK&Recreational and sporting activities\\
145&ZAXM&Cultural services\\
146&ZAXS&All restaurants and hotels\\
147&ZAXU&All catering services\\
148&ZAXW&Restaurants, cafes etc\\
149&ZAYC&Canteens\\
150&ZAYE&Accommodation services\\
151&ZAYO&Social protection\\
152&ZAYQ&All insurance\\
153&ZAYS&Insurance connected with the dwelling\\
154&ZAYU&Insurance connected with health\\
155&ZAYW&Insurance connected with transport\\
156&ZAZA&All financial services n.e.c.\\
157&ZAZC&All financial services other than FISIM\\
158&ZAZE&Other services n.e.c.\\
159&ZWUT&Education\\ \hline
\multicolumn{3}{c}{ \textbf{Deflators}}\\ \hline
160&ABJS&Consumption\\
161&FRAH&RPI: Total Food\\
162&ROYJ&Wages\\
163&YBGB&GDP Deflator\\ \hline
\multicolumn{3}{c}{ \textbf{Money Series}}\\ \hline
164&M4ISA&M4 Deposits PNFCs\\
165&M4OSA&M4 Deposits OFCs\\
166&M4PSA&M4 Deposits Households\\
167&MALISA&M4 Lending Total\\
168&MALOSA&M4 Lending PNFCs\\
169&MALPSA&M4 Lending Households\\ \hline
\multicolumn{3}{c}{ \textbf{Asset Prices}}\\ \hline
170&&Real nationwide house prices\\
171&GDF Data&FTSE All Share Index\\
172&IMF Data&Nominal Effective Exchange Rate (NEER)\\
173&GDF Data&Pounds to Euro\\
174&GDF Data&Pounds to US dollar\\
175&GDF Data&Pounds to Canadian dollar\\
176&GDF Data&Pounds to Australian dollar\\
\hline
\end{longtable}

\newpage
\bibliographystyle{authordate1}
\bibliography{Bibliography}
\end{document}